\documentclass{mn2e}

\usepackage{psfig,subfigure,graphicx,mncite}

\newcommand{\bvec}[1]{\mbox{\boldmath ${#1}$}}

\begin{document}

\title[X-ray emission from T Tauri stars]
{X-ray emission from T Tauri stars}

\author
[M. Jardine, A. Collier Cameron, J.-F. Donati, S.G. Gregory \& K. Wood]
{M. Jardine$^1$, \thanks{E-mail: moira.jardine@st-and.ac.uk}
A. Collier Cameron$^1$, J.-F. Donati$^2$, S.G. Gregory$^1$ \& K. Wood $^1$
\\
$^1$SUPA, School of Physics and Astronomy, Univ.\ of St~Andrews, 
St~Andrews, 
Scotland KY16 9SS \\
$^2$Laboratoire dÕAstrophysique, Observatoire Midi-Pyr\'en\'ees, 14 
Av. E. Belin, F-31400 Toulouse, France \\
\\
} 

\date{Received; accepted 2005}

\maketitle

\begin{abstract}

We have modelled the X-ray emission of T Tauri stars assuming that they have isothermal, magnetically-confined coronae. These coronae extend outwards until either the pressure of the hot coronal gas overcomes the magnetic field, or, if the corona interacts with a disk before this happens, by the action of the disk itself. This work is motivated by the results of the Chandra Orion Ultradeep Project (COUP) that show an increase in the X-ray emission measure with increasing stellar mass. We find that this variation (and its large scatter) result naturally from the variation in the sizes of the stellar coronae. The reduction in the magnitude of the X-ray emission due to the presence of a disk stripping the outer parts of the stellar corona is most pronounced for the lower mass stars. The higher mass stars with their greater surface gravities have coronae than typically do not extend out as far as the inner edge of the disk and so are less affected by it. For these stars, accretion takes place along open field lines that connect to the disk. By extrapolating surface magnetograms of young main sequence stars we have examined the effect on the X-ray emission of a realistic degree of field complexity. We find that the complex fields (which are more compact) give densities of some $(2.5 - 0.6) \times 10^{10}$cm$^{-3}$. This is consistent with density estimates of $(1-8)\times 10^{10}$cm$^{-3}$ from modelling of individual flares. A simple dipole field in contrast gives densities typically an order of magnitude less. For the complex fields, we also find surface hotspots at a range of latitudes and longitudes with surface filling factors of only a few percent. We find that the dipolar fields give a relationship between X-ray emission measure and stellar mass that is somewhat steeper than observed, while the complex fields give a relation that is shallower than observed. This may suggest that T Tauri stars have coronal fields that are slightly more extended than their main sequence counterparts, but not as extended as a purely dipolar fields.

\end{abstract}


\section{Introduction}

The recent Chandra Orion Ultradeep Project (COUP) has provided a remarkable dataset of the X-ray emission of pre-main sequence stars. It shows that for these T Tauri stars (as for main sequence field stars) the X-ray emission measure rises slowly with stellar mass. For the T Tauri stars that show signs of active accretion, however, the correlation is less well defined \cite{preibisch_COUP_insights_05}. These COUP stars have rotation rates (or Rossby numbers) that place them in the {\em saturated} or {\em supersaturated} part of the X-ray emission/rotation rate relation, but with emission measures slightly below that of their main-sequence counterparts. This suppression of the X-ray emission also appears to be related to the presence of active accretion from the surrounding disk. The detection of significant rotational modulation in some of the saturated stars suggests that the saturation of the X-ray emission is not the result of the complete coverage of the stellar surface in emitting regions \cite{flaccomio_COUP_rotmod_05}. Rather, the emitting regions must be distributed unevenly around the star and they do not extend significantly more than a stellar radius above the surface. 

Observations of flares also yield information about the structure of T Tauri coronae. Although flares typical of the compact flares seen on main-sequence stars have been observed in the COUP sample, extremely large flares have also been detected. Estimates of plasma densities based on these flare observations are in the range of $(1-8) \times 10^{10}$cm$^{-3}$ \cite{favata_COUP_flares_05}. The derived lengths of the flaring loops are around the location of the co-rotation radius (~5R$_\star$) for the most reliable estimates, although some loop lengths appear to much greater, suggesting perhaps that these flares have taken place in loops that link the star and the surrounding disk.

 \begin{figure}
  \begin{center}
   \includegraphics[width=2.5in]{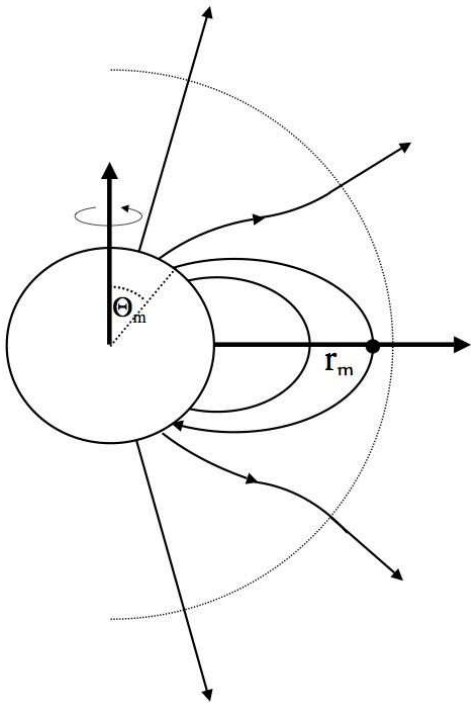}
   \caption{Magnetic structure of a dipole field that has been distorted by the outward pressure of the hot coronal gas trapped on the field lines. Beyond some radius (shown dotted) the pressure of the hot coronal gas is large enough to force open the magnetic field lines. An example is shown of the {\em last closed field line} that intersects the equatorial plane at a radial distance $r_m$ and emerges from the stellar surface at a co-latitude $\Theta_m$. If the coronal temperature and hence pressure is increased, or the stellar field strength is decreased, then $r_m$ moves inwards and the angle $\Theta_m$ increases. As a result, the fraction of the stellar surface that is covered in closed field capable of confining X-ray emitting gas decreases.}
   \label{dipole_cartoon}
  \end{center}
\end{figure}

The X-ray data from COUP therefore suggest that T Tauri stars have coronae that are structured on a variety of length scales, ranging from compact loops on scales of a stellar radius to much longer structures on the scale of the inner regions of an accretion disk. Observations of T Tauri magnetic fields also point to a very complex field structure, with localised regions of high field strength on the stellar surface. Unpolarized Zeeman-splitting measurements for photospheric lines with high Lande g-factors typically yield field strengths on T Tauri stars of order 2 to 3 kG \cite{johns_krull_BPTau_99,johns_krull_NTT_04}, with surface filling factors of order some tens of percent. Similarly, the observed splittings of the circularly-polarised Zeeman $\sigma$ components in the He I 5876 line indicate that the accretion streams impact the stellar surface in regions where the strength of the  local magnetic field is of order 3 to 4.5 kG \cite{guenther_TTS_99,johns_krull_BPTau_99,symington_tts_05}. As \scite{valenti_johns_krull_04} point out, unpolarised Zeeman-broadening measurements are sensitive to the distribution of magnetic field strengths over the stellar surface, but yield little spatial information. Circular polarization methods such as Zeeman-Doppler imaging give better spatial information about the magnetic polarity distribution, but have limited ability to determine field strengths at the level of the photosphere.

The rate at which the field strength  declines with radial distance from the star depends on the complexity of the magnetic polarity distribution on the stellar surface. On the Sun, much of the flux emerging from small-scale bipolar regions in the photosphere connects locally to nearby regions of opposite polarity. The locally-connected field lines do not contribute to the coronal field at heights greater than the footpoint separation. As a result, the power-law dependence of the average coronal field on height steepens with increasing field complexity. A simple model assuming a global dipole with a surface field strength of order 2 to 3 kG would probably assume too high a value for the surface field, and allow it to decline too slowly with radial distance.

Zeeman-Doppler images of post-T-Tauri stars give us a much better idea of the complexity of the magnetic field, and hence of the rate at which it falls off with height. The surface resolution of a Zeeman-Doppler image is determined by the ratio of the star's rotational line broadening to that of the intrinsic line profile convolved with the instrumental resolution. On length scales smaller than a resolution element, the circular polarisation signals from small-scale field concentrations of opposite polarity cancel. The Zeeman broadenings seen in unpolarised light, however, do not. The ``field strengths" in Zeeman-Doppler images are in effect magnetic flux densities, averaged over a surface resolution element \cite{donati97abdor95}. For this reason, the highest flux densities seen in Zeeman-Doppler images are always less than the maximum field strengths determined from unpolarised Zeeman splittings. Potential-field extrapolations from Zeeman-Doppler images should give a realistic measure of the field strength as a function of height beyond a few tenths of a stellar radius above the photosphere. 

The X-ray coronae of these pre-main sequence stars appear therefore to display a degree of spatial complexity and activity that is similar to that of their very active (saturated or supersaturated) main sequence counterparts, but with some significant differences. These differences (the suppression of X-ray emission and the greater scatter in the $L_x - M_\star$ relation) appear to be related to the presence of active accretion. For main sequence stars  the observed variation with rotation rate of both the magnitude of the X-ray emission its rotational modulation can be explained by the centrifugal stripping of the outer parts of the corona \cite{unruh97loops,jardine99stripping,jardine04stripping}. This becomes important once the star is rotating sufficiently rapidly that its co-rotation radius is within the corona. \scite{ivanova_taam_braking_03} have shown that this process can also reduce the angular momentum loss rate of rapid rotators (relative to more slowly rotating stars) without requiring a saturation of dynamo activity at high rotation rates. While most T Tauri stars are not rapidly rotating, their lower surface gravities and therefore greater pressure scale heights mean that their coronae may be more extended than their main-sequence counterparts. This suggests that a similar process may be important for these stars too. The presence of active accretion from the inner edge on an accretion disk may distort the corona of a T Tauri star, leading to a similar (but additional) stripping effect. 

The aim of this paper is to model the X-ray emission from T Tauri stars, using the data (masses, radii, rotation periods, X-ray emission measures and temperatures) provided by the COUP results \cite{getman_COUP_list_05a,getman_COUP_list_05b}. We show these apparently puzzling features of X-ray emission from T Tauri stars can be explained in the context of the varying sizes of stellar coronae that are limited either by  the gas pressure of the stellar corona or by the presence of a disk.
 \begin{figure}
 \begin{center}
   \includegraphics[width=3in]{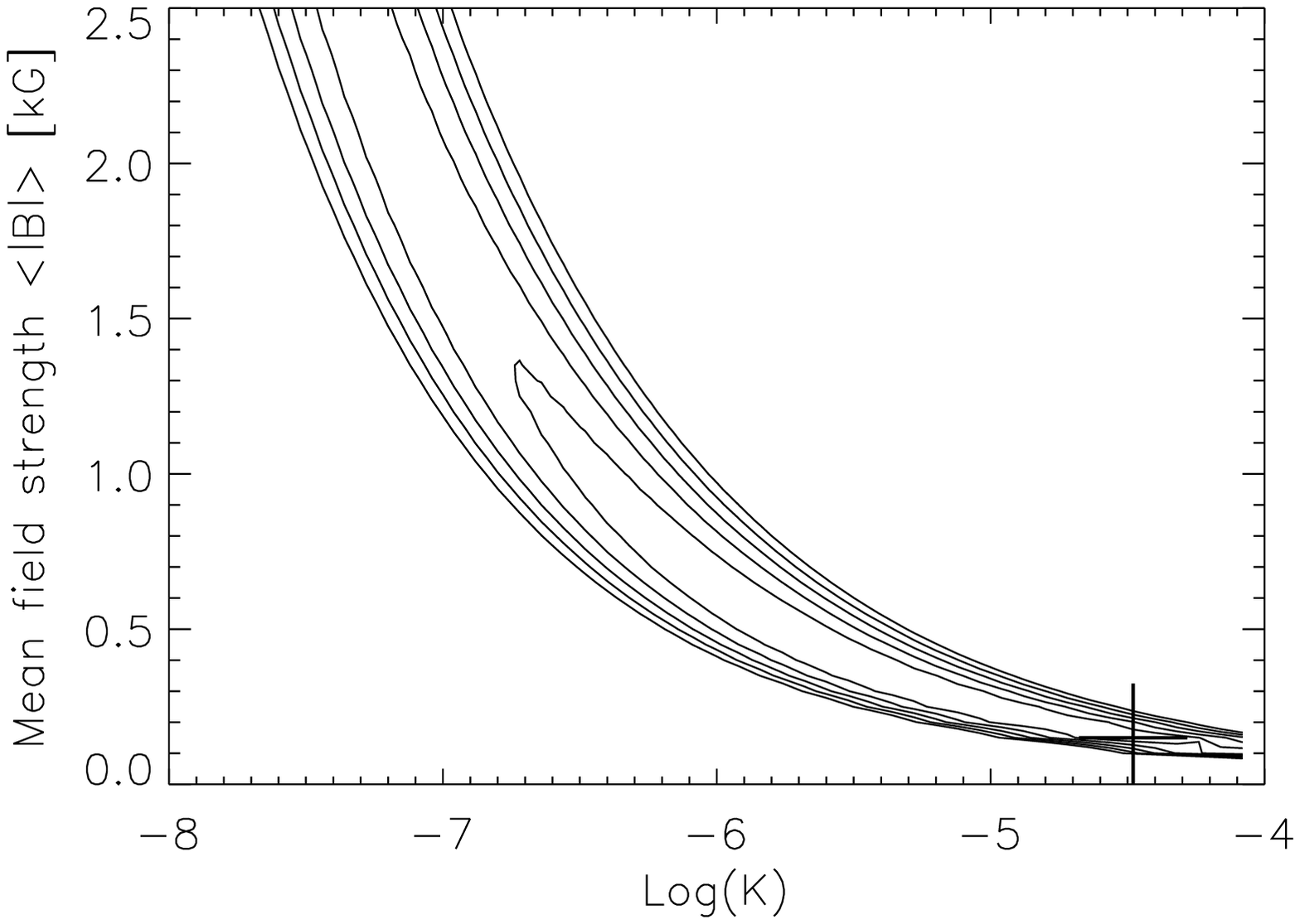}
  \includegraphics[width=3in]{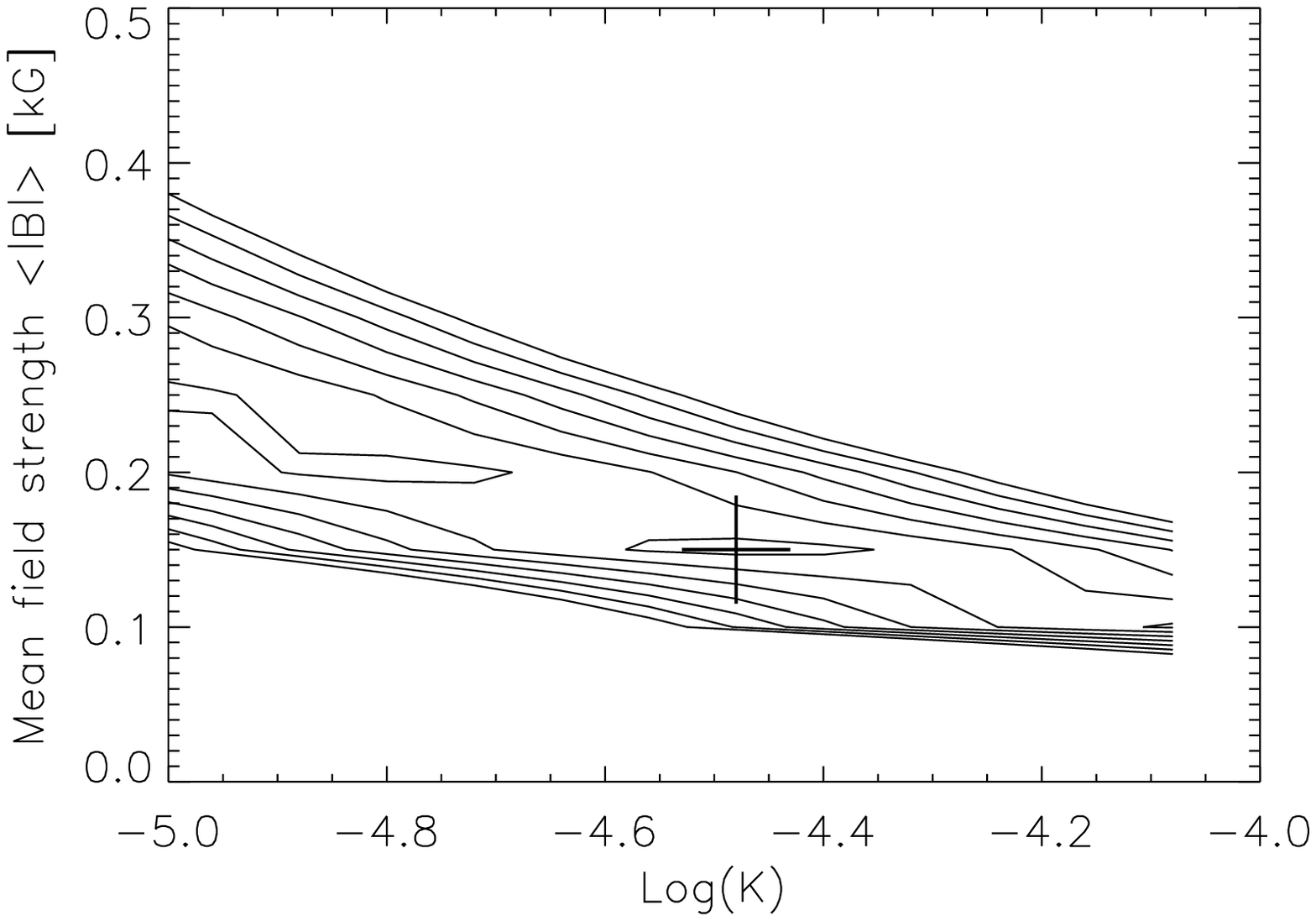}
   \caption{Contours of D (a measure of the goodness of fit to the COUP emission measures) as a  function of the average value of the magnitude of the surface field strength and the constant of proportionality K which determines the pressure $p_0$ at the base of the corona since $p_0 = K B_0^2$. The cross marks the position of the minimum value of D and the lower panel shows in greater detail the area around this minimum. The plot shown is for models of the higher-temperature emission measures. A plot for the lower-temperature emission measures is qualitatively similar.}
   \label{chisq_contours}
  \end{center}
\end{figure}

\section{Simple model}
\subsection{Passive disk}
 \begin{figure*}
  \begin{center}
   \includegraphics[width=3in]{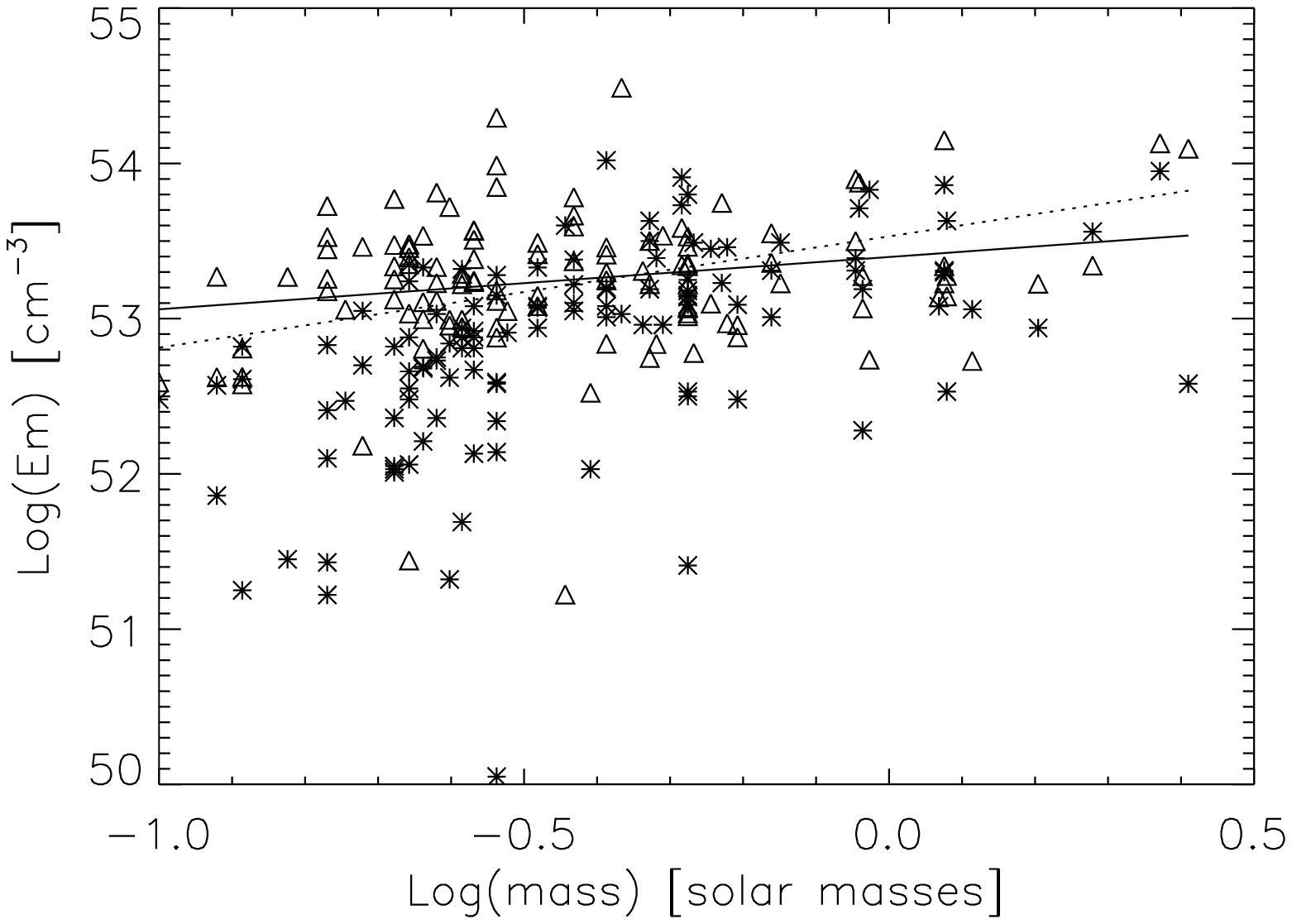}
   \includegraphics[width=3in]{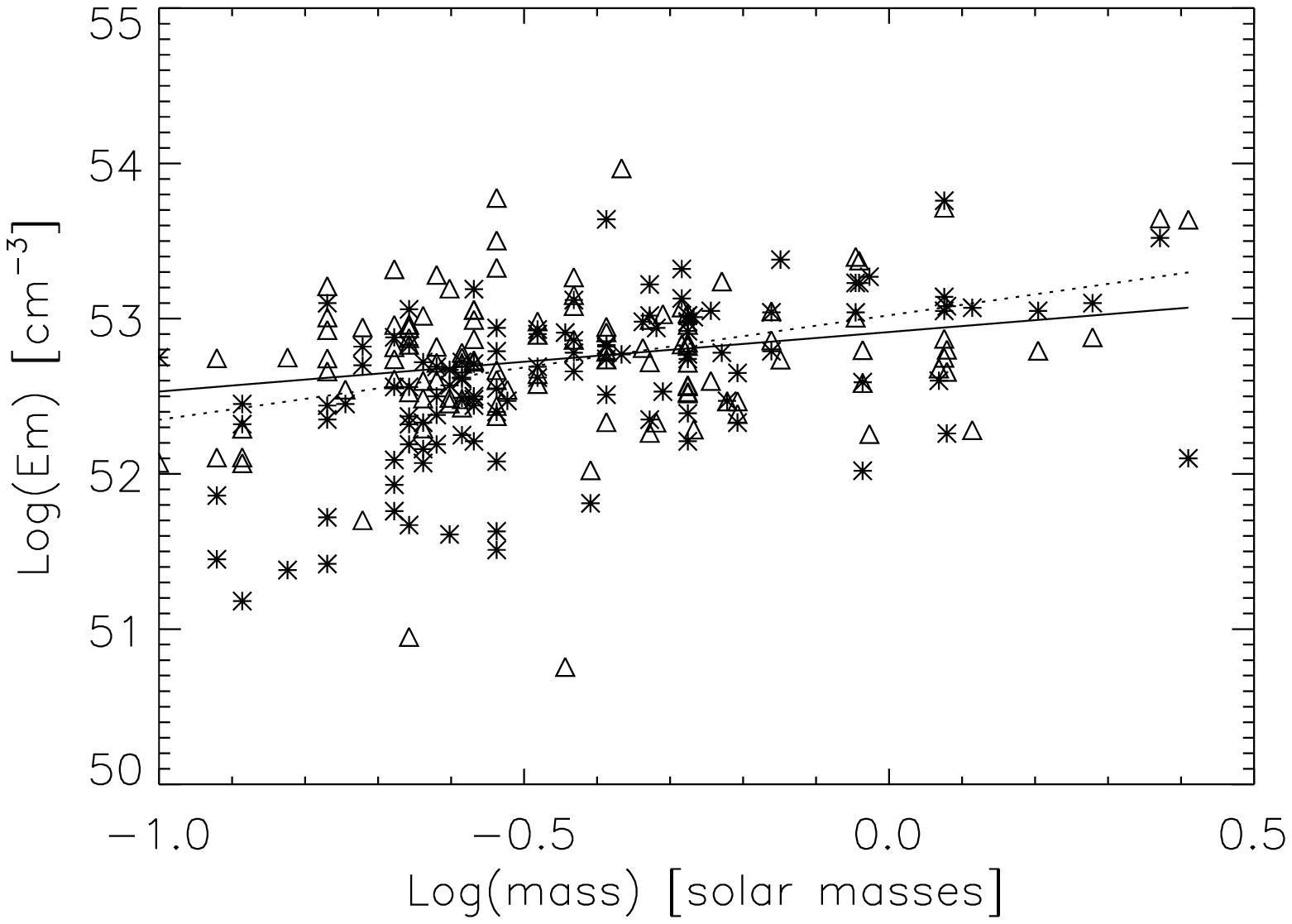}
   \caption{Emission measures for the high temperature (left) and low temperature (right) ranges from the COUP sample. Triangles show the observed values while crosses show the values calculated for a dipole field with values of the base magnetic and plasma pressures that give the best fit to the data. For each star we assume that any disk that may be present is {\em passive} in the sense that the maximum extent of the corona is limited only by the ability of the magnetic field to contain the hot coronal gas. The solid line is the best fit to the calculated values, while the dotted line is the best fit to the observed values, assuming the errors quoted for the log(emission measures).}
   \label{data_dip_full}
 \end{center}
\end{figure*}
 \begin{figure*}
  \begin{center}
     \includegraphics[width=3in]{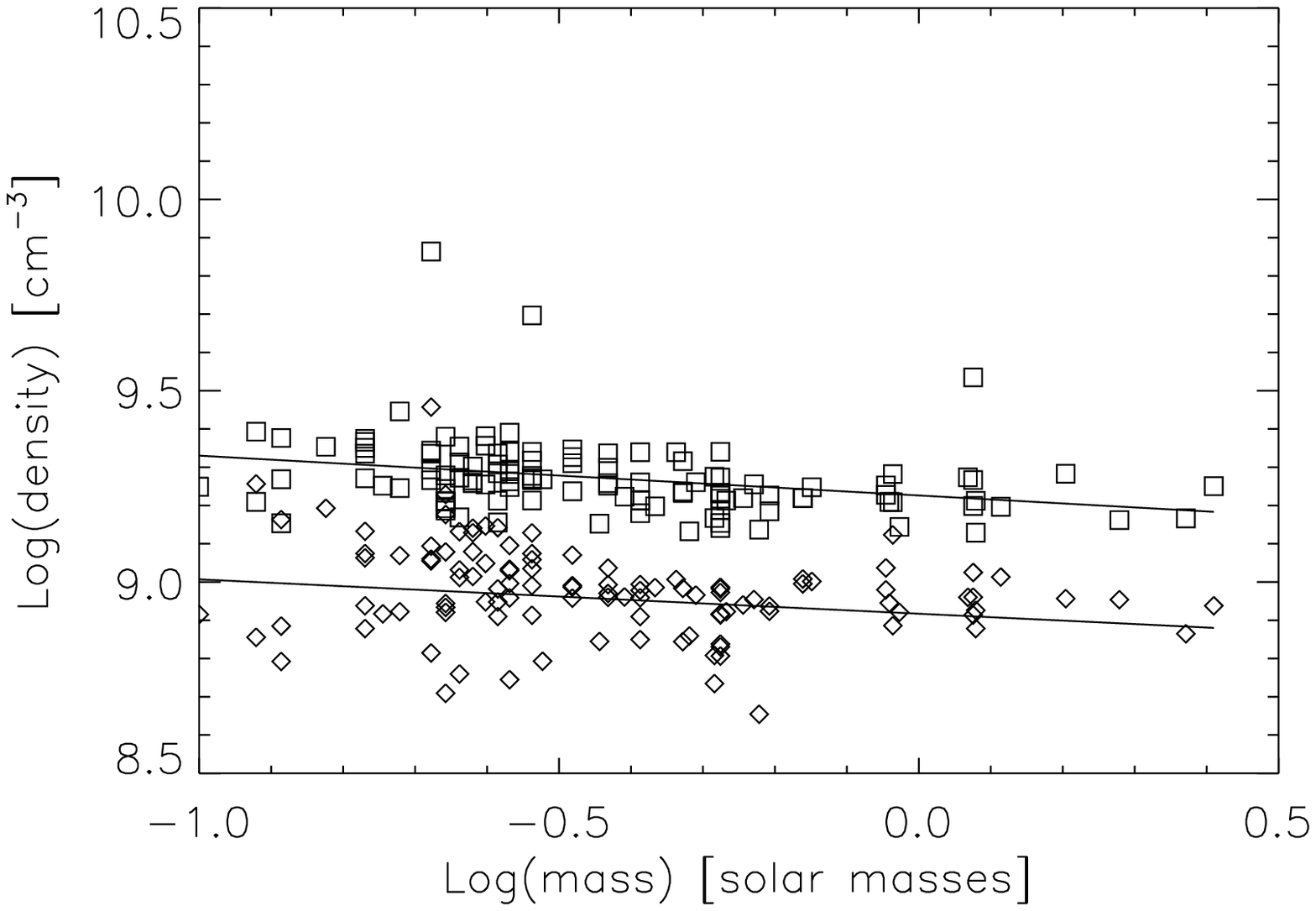}
     \includegraphics[width=3in]{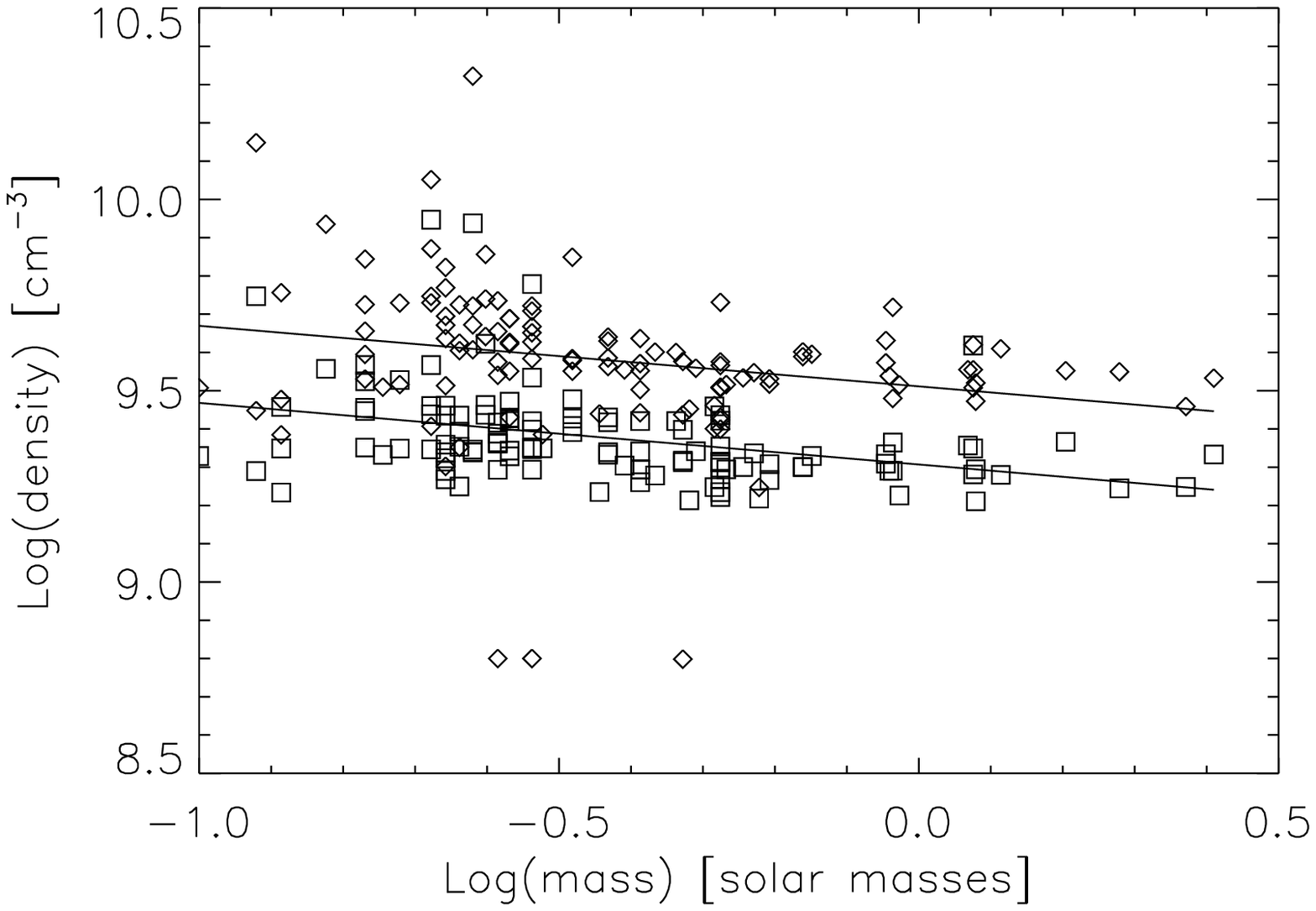}
    \caption{Calculated emission-measure weighted densities for both the high temperature range (diamonds) and low temperature range (squares). In the left panel it has been assumed that any disks that may be present are {\em passive} so that the size of the corona is limited only by the ability of the magnetic field to contain the hot coronal gas. The right panel, in contrast, shows the densities obtained by assuming that the same stars each have instead an {\em active disk} so that the extent of the corona is limited by the action of the disk truncating the corona at the co-rotation radius. The lines show the best fits to the calculated densities.}
   \label{dens_dip}
  \end{center}
\end{figure*}

We begin by addressing the question: what is the extent of the star's corona? We consider first the situation where each star has a disk that is {\em passive} in the sense that it does not distort the star's magnetic field. The coronal extent can then be estimated by calculating the height at which the gas pressure begins to exceed the magnetic pressure. In order to do this however we need a model for the star's magnetic field structure. As a simple example, we take a dipolar field geometry (see Fig. \ref{dipole_cartoon}) for which
\begin{eqnarray}
B_r  & =  & \frac{2m\cos\theta}{r^3} \\
B_\theta & = & \frac{m\sin\theta}{r^3}.
\end{eqnarray}
If we assume that the gas that is trapped on each closed field line is isothermal and in  hydrostatic equilibrium then the gas pressure is $p=p_{0}e^{\frac{m}{kT}\int g_{s}ds}$ where $g_{s} =( {\bf g.B})/|{\bf B}|$ is the component of
gravity along the field. Here, for a star with rotation rate $\omega$,
\begin{equation}
 g(r,\theta) = \left( -GM_{\star}/r^{2} + 
                     \omega^{2}r\sin^{2} \theta,
		     \omega^{2}r\sin \theta \cos\theta 
             \right).
 \label{grav}   
\end{equation}
 
 The gas pressure at the footpoint of the field line $p_0$ is a free parameter of this model. For this initial simple analysis we choose to scale this to the magnetic pressure of the average surface field strength $B_0 = < |B|>$ such that $p_{0}=K B^{2}_{0}$ where $K$ is a constant that is the same on every field line. By scaling $K$ up or down we can scale the overall level of the coronal gas pressure and hence the density and emission measure (see Appendix A for the calculation of the emission measure and the density). For each value of K we can also determine the variation of the gas pressure along each field line and hence the height beyond which the gas pressure exceeds the magnetic pressure. We show in Fig. \ref{dipole_cartoon} the {\em last closed field line}, i.e. the most extended field line that can still contain coronal gas. Since the flux function
\begin{equation}
A = \frac{\sin^2\theta}{r}
\label{flux_function}
\end{equation}
is constant along field lines, if this field line crosses the equator at $r=r_m$, then it has a footpoint on the stellar surface at a co-latitude $\Theta_m$ given by
\begin{equation}
\sin^2\Theta_m = \frac{r_\star}{r_m}.
\end{equation}
If the field strength is decreased, or the temperature is increased, then the extent of the last closed field line will decrease and so $\Theta_m$ will increase.
 \begin{table}
\caption{This shows, for a dipole field, the  values of the average field strength $B_0 = <|B|>$ at the base of the corona and the corresponding gas pressure $p_0 = KB_0^2$  that give the best fit to the observed emission measures from the COUP sample. By fitting relationships of the form $EM \propto M_\star^a$ and $\bar{n}_e \propto M_\star^b$ to the calculated values of the emission measures and densities we have determined the values of $a$ and $b$.}
\begin{center}
\begin{tabular}{|c|c|c|c|c|}
\hline
                                   & Passive disk &            &   Active  disk     &    \\
\hline
  T                                        &  High         & Low            &   High              & Low  \\
\hline
 $<|B|> [G]$                         &  100         & 50               &   150               &  50 \\
 \hline
   $p_0$ [dyne cm$^{-2}$]   &   1.9          & 1.2              &   7.5                & 1.4     \\
 \hline
 a                                         & 0.34           & 0.38           &  0.70               & 0.62     \\
 \hline
 b                                        & -0.09           & -0.10          &-0.16                & -0.16   \\
 \hline

\end{tabular}
\end{center}
\label{table1}
\end{table}
Our model now has six parameters: the mass, radius and rotation rate of the star, the coronal temperature and the values of the magnetic and gas pressures at the base of the corona. For a large number of stars in the COUP sample the masses, radii and rotation rates have been determined. A two-temperature fit to the X-ray spectra then gives emission measures at two different temperatures. This leaves only the  base gas and magnetic pressures undetermined. The value of $p_0$ is simply a scaling factor that raises or lowers the overall coronal pressure. We have chosen to make it proportional to the magnetic pressure since this gives a good agreement between predicted and observed values for emission measures and mean densities both the the Sun and other main-sequence stars \cite{jardine02structure,jardine02xray}. We can then simply ask which combination of $(p_0,B_0^2)$ gives the best fit to the emission measures of the stars in the COUP sample. To do this, we select the 116 stars from the COUP sample for which there are stellar radii, masses and rotation periods \cite{getman_COUP_list_05a,getman_COUP_list_05b}. We exclude one star for which the apparent co-rotation radius is less than the stellar radius.  For each star we then determine (for each of the two coronal temperatures quoted in the COUP sample) the emission measure (EM) and calculate, for the $N$ stars
\begin{equation}
  D = \sum_{i=1}^{i=N} \frac{(\log_{10}(EM_{\rm obs,i} )- \log_{10}(EM_{\rm calc,i}))^2}{\sigma_i^2}
  \label{chisq}
\end{equation}
where the $\sigma_i$ are the errors in the $\log_{10}(EM)$ values. We show in Fig. \ref{chisq_contours} contours of D for a range of values of $(p_0,B_0^2)$. Clearly, as we choose higher values of the base field strength (which will tend to give a larger corona) we need to have smaller values of the pressure (and hence the density) to give the same emission measure.  Table \ref{table1} shows the values of  $(p_0,B_0^2)$ that give the best fit to the data and Fig. \ref{data_dip_full} shows both the observed and calculated emission measure for these values. Even this simple model reproduces reasonably well the variation of emission measure with stellar mass and the degree of scatter at each value of $M_\star$. For every value of $M_\star$ there are of course many possible combinations of the other parameters ($R_\star, P_{\rm rot}, T$) that determine the extent of the corona and the emission measure.  If we fit a relationship of the form $EM \propto M_\star^a$ to the calculated emission measures we find that the slopes in both the high and low temperature ranges are shallower than for a corresponding fit to the data, which are respectively 0.68 and 0.65 (assuming the errors quoted for the log(emission measures)). We remind the reader at this point that we are using only a subset of the stars  from the COUP survey. We have restricted ourselves only to those that, in addition to emission measures and coronal temperatures, also have published masses, radii and rotation periods. For this subset of stars, the slope of the observed emission measure - mass relation is shallower than we would find if we were to use the full sample of emission measures.

We can also (for each of the two temperatures quoted in the COUP database) calculate the emission measure-weighted density
 \begin{equation}
 \bar{n}_e =\frac{ \int n_e^3 dV}{ \int n_e^2 dV}
 \end{equation}
(see Fig. \ref{dens_dip}). If we fit a relationship of the form $\bar{n}_e \propto M_\star^b$ to the calculated emission-measure weighted densities, we find that the slopes at the two temperature ranges are similar, but the magnitude of the density for the lower temperature range is systematically lower. As shown in the left-hand panel of Fig. \ref{dens_dip}, these are rather lower than the values of $(1-8)\times 10^{10}$cm$^{-3}$ determined by \scite{favata_COUP_flares_05} from modelling flares observed in the COUP study, and indeed they are also lower than the values of $10^{10} - 10^{13}$cm$^{-3}$ calculated for young main sequence stars (for a comprehensive review see \scite{guedel_review_04}). 
 \begin{figure}
   \includegraphics[width=3.0in]{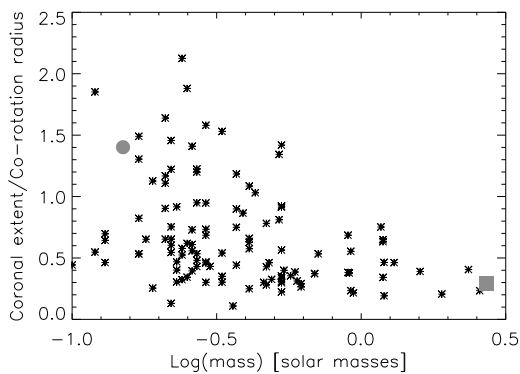}
   \caption{Coronal extent (in units of the Keplerian co-rotation radius) as a function of stellar mass for stars in the COUP sample. Two stars (marked with a circle and square) have been selected as examples. Their coronal structure is shown more fully in Figs. \ref{field_geom_mxd180} and \ref{field_geom_lqhya}.}
   \label{coronal_extent}
\end{figure}

Fig. \ref{coronal_extent} shows the coronal extents (in units of the co-rotation radius for each star) that gave this best-fit value of $(p_0,B_0^2)$.  The more massive stars typically have coronae whose extents (relative to their co-rotation radii) are smaller than that of the lower mass stars. Indeed, for these field strengths and pressures many of the higher mass stars have coronae that do not extend as far as the co-rotation radius. It is the lower-mass stars with their lower surface gravity that are more likely to have coronae that extend out beyond the co-rotation radius and hence to suffer more interaction with any disks that may be present. This may have significant implications for the nature of the interaction between the stellar field and the disk and the exchange of torques between them, since for these higher mass stars it may well be that the disk intercepts not the closed (dipolar) field of the star, but the open field of its wind.
 \begin{figure*}
 \begin{center}
   \includegraphics[width=3in]{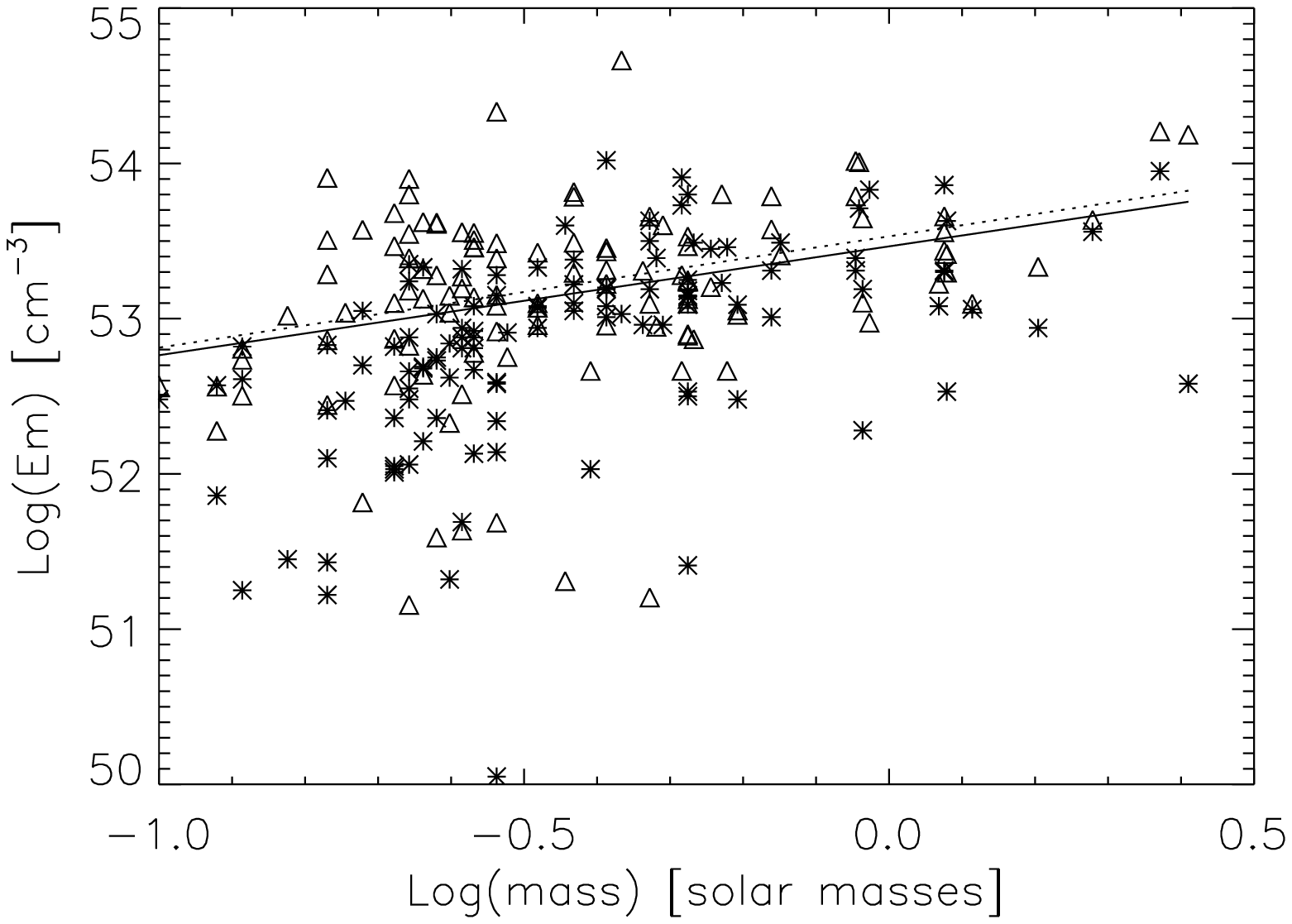}
   \includegraphics[width=3in]{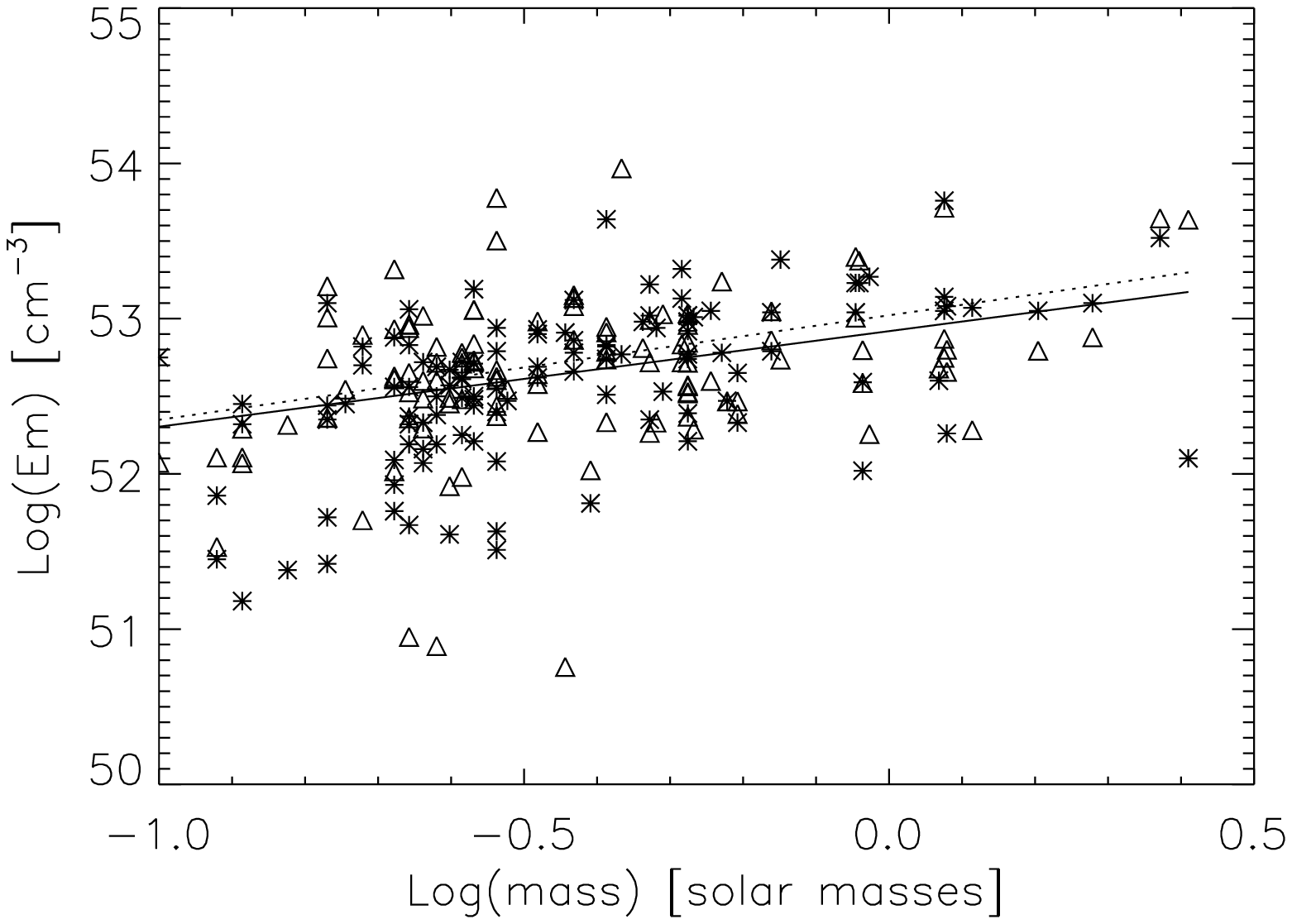}
   \caption{Emission measures for the high temperature (left) and low temperature (right) ranges from the COUP sample. Triangles show the observed values while crosses show the values calculated for a dipole field with values of the base magnetic and plasma pressures that give the best fit to the data. We assume that each star has an {\em active} disk in the sense that the maximum extent of the corona is limited by the inner edge of the disk which is assumed to be at the co-rotation radius. The solid line is the best fit to the calculated values, while the dotted line is the best fit to the observed values, assuming the errors quoted for the log(emission measures).}
   \label{dataplusmodel}
 \end{center}
\end{figure*}

 \subsection{Active disk}
 Up to this point, we have considered only the situation where each star has a  {\em passive} disk. Each star's corona then extends out until the pressure of the coronal gas exceeds that of the magnetic field confining it, causing the field to open up and release the gas into the stellar wind. For the lower-mass stars, however, whose coronae extend to distances greater than the co-rotation radius, the presence of a disk extending in as far as the co-rotation radius may disrupt this field. If this disruption is extreme, the presence of a disk may alter the structure of the stellar corona by shearing those field lines that pass through it, causing them to become open and therefore allowing coronal gas to escape in a  wind \cite{lynden_bell_boily_94,lovelace_inflate_95}. We refer to this as an {\em active} disk. This process is likely to reduce the overall X-ray emission measure since it converts closed field lines that could contain hot X-ray emitting coronal gas into wind-bearing open field lines that will be dark in X-rays. This effectively places an upper limit on the size of the corona, since closed field lines cannot extend beyond the co-rotation radius. This would have the greatest impact on the lower-mass stars whose coronae are naturally more extended because of their lower surface gravity. The X-ray emission from stars whose coronae did not extend as far as the co-rotation radius would be affected much less severely.
 
We therefore take the extent of the corona to have a maximum value of the location of the inner edge of the disk which we take to be at the co-rotation radius We then calculate the emission measure of the gas contained in each stellar corona for a range of values of both the mean magnetic field at the stellar surface and the base pressure of the corona.  We show in Table \ref{table1} the values of the base gas and magnetic pressure that give the  minimum value of D (as defined by (\ref{chisq})) and show in Fig. \ref{dataplusmodel} both the observed and calculated emission measures at these best-fit values. The presence of an active disk clearly gives a rather better fit to the emission measures at both temperature ranges. This is mainly because of the lower mass stars which have coronae that extend beyond the co-rotation radius and whose emission measure is suppressed by the action of the disk stripping the corona. As a result, these stars have a lower emission measure than would be produced with a passive disk and so the slope of the fitted line is larger and closer to that found for the observed emission measures. While the calculated densities (as shown in Fig. \ref{dens_dip}) are still lower than observed, they are a little higher than if the disk is assumed to be passive and the slope of their variation with mass is greater. This is because the best-fit value of the base pressure (and hence the base density) is higher for the an active disk where the typically smaller sizes of the stellar coronae require higher densities to match the observed emission measures. The lower-temperature coronae with the smaller scale heights are less affected by the presence of the disk and their density typically increases by less when an active disk is present.

 \subsection{Surface hot spots}
\begin{figure}
   \includegraphics[width=3in]{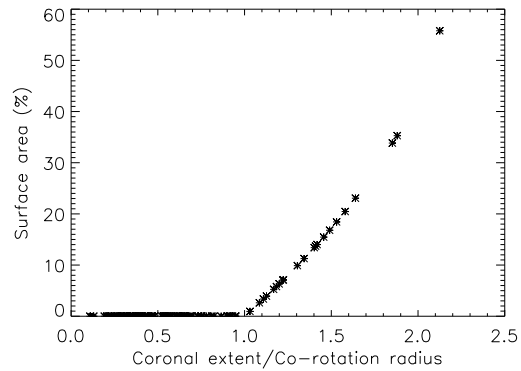}
   \caption{Crosses show the upper limit to the fraction of the stellar surface area that is covered in field lines that pass through the disk. A dipole field has been assumed.}
   \label{ff_dip}
\end{figure}

 Observations of magnetic fields on T Tauri stars and modelling of accretion signatures suggest that the fraction of the surface area of the star that is covered in accreting hot spots is very small, of the order of a few percent \cite{muzerolle_mdot_03,calvet_highmass_04,muzerolle_vlm_05,valenti_johns_krull_04,symington_tts_05}. We can place an upper limit on this by determining which field lines (in a corona whose size is limited only by pressure balance) would intersect the disk. Some subset of these field lines will be capable of supporting an accretion flow. We can determine this surface area fraction $F$ by calculating the area of the two annuli on the stellar surface that lie between $\Theta_{\rm m}$ (the footpoint of the last closed field line) and $\Theta_{KCR}$ (the footpoint of the field line that passes through the co-rotation radius). As a fraction of the total stellar surface area this gives
 \begin{equation}
 F = \cos \Theta_{\rm m} - \cos \Theta_{\rm KCR}.
 \end{equation}
 This fraction is shown in Fig. \ref{ff_dip} as a function of the coronal extent. If the corona does not extend out as far as the inner edge of the disk, then only those open field lines that pass close to the equator will intersect the disk. Since these are likely to come from the polar regions, their area coverage at the stellar surface is very small. For those coronae that do extend beyond the inner edge of the disk, the area of the surface covered by those field lines that cross the disk depends on both the value of the co-rotation radius and also the maximum extent of the corona as determined by pressure balance. This can be seen from (\ref{flux_function}) since, if the stellar radius is sufficiently small compared to either the co-rotation radius or the maximum extent ($r_m$) of the corona,
 \begin{equation}
 F \approx \frac{r_\star}{r_{\rm KCR}} - \frac{r_\star}{r_{\rm m}}.
 \end{equation}
 This suggests that it is the lower-mass stars with their more extended coronae that should show the largest hotspots. 

 \begin{figure*}
  \begin{center}
   \includegraphics[width=7in]{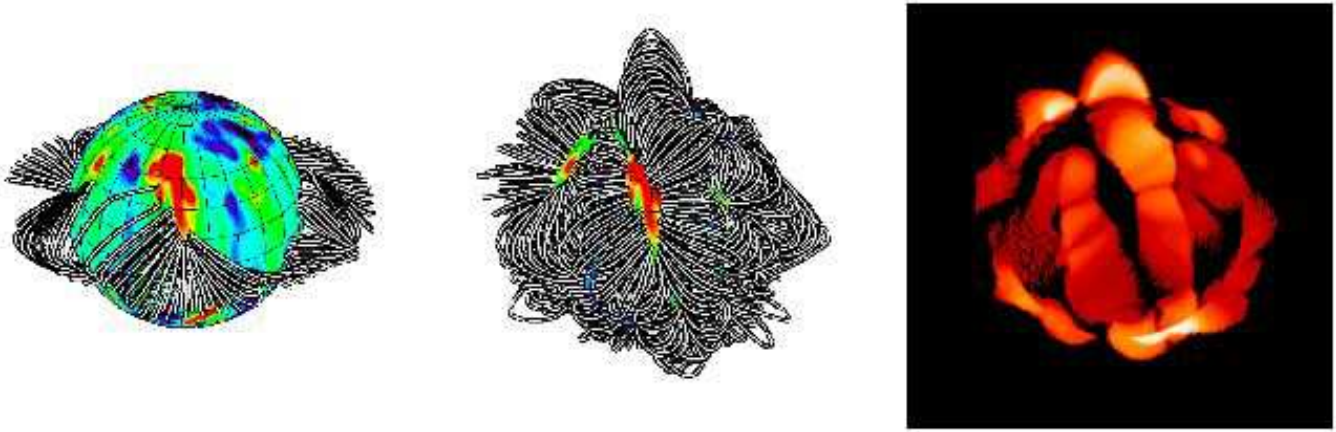} 
   \caption{Calculated coronal structure based on an example surface magnetic field distribution for one of the lower mass stars (shown as a circle in  Fig. \ref{coronal_extent} ) where it is the presence of the disk that is primarily limiting the extent of the corona. In this example the stellar mass is 0.15 M$_\odot$, the radius is 4.02 R$_\odot$, the rotation period is 17.91 days and the natural extent of the corona (in the absence of a disk) would be at most 1.58 times the co-rotation radius. The left panel shows the field lines that could support an accretion flow, drawn from the inner edge of the accretion disk which is situated at the co-rotation radius. The middle panel shows examples of the field lines that are closed and contain X-ray emitting gas and  the right panel shows the corresponding structure of the X-ray corona. }
   \label{field_geom_mxd180}
 \end{center}
\end{figure*}
 \begin{figure*}
  \begin{center}
     \includegraphics[width=7in]{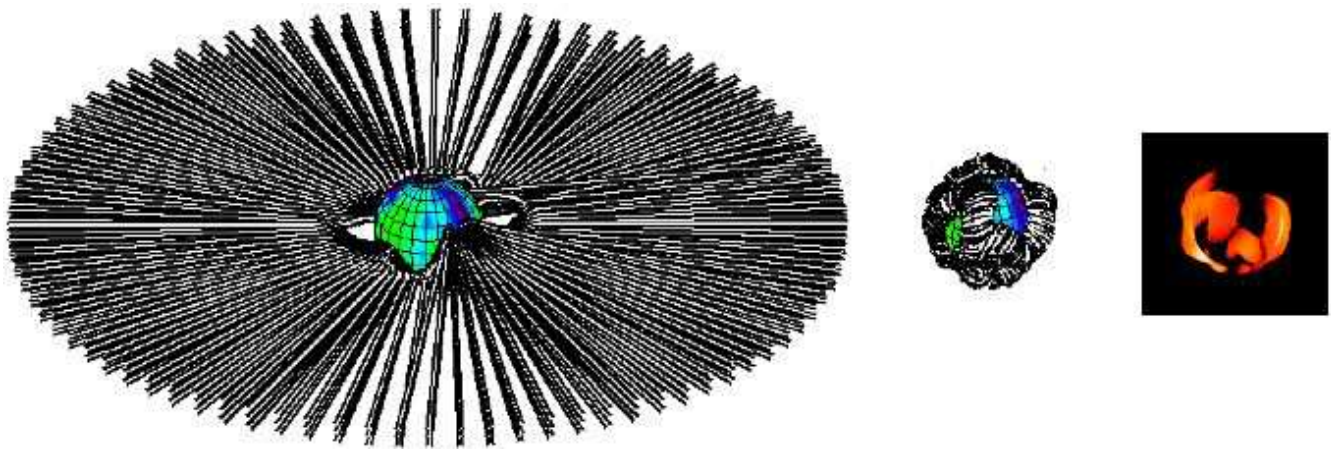} 
   \caption{Calculated coronal structure based on an example surface magnetic field distribution for one of the higher mass stars (shown as a square in  Fig. \ref{coronal_extent} ) where it is the pressure of the hot coronal gas that is primarily limiting the extent of the corona. In this example the stellar mass is 2.57 M$_\odot$ the radius is 1.74 R$_\odot$, the rotation period is 1.38 days and the natural extent of the corona (in the absence of a disk) would be at most only 0.26 times the co-rotation radius. The left panel shows the field lines that could support an accretion flow, drawn from the inner edge of the accretion disk which is situated at the co-rotation radius. The middle panel shows examples of the field lines that are closed and contain X-ray emitting gas and  the right panel shows the corresponding structure of the X-ray corona.}
   \label{field_geom_lqhya}
 \end{center}
\end{figure*}

 It seems unlikely however that T Tauri stars will have simple aligned dipolar fields. The observations of significant rotational modulation of the X-ray emission from many of the stars in the COUP sample are a clear indication that their fields are more complex than a simple dipole. The predominance of a period for the X-ray emission that is equal to (or half) that of the photometric period suggests the presence of either one dominant region of X-ray emission on the stellar surface or two at opposite longitudes. This is very similar to the predictions of the X-ray rotational modulation for AB Dor based on extrapolation of surface fields determined from Zeeman-Doppler imaging \cite{jardine02structure,jardine02xray}. These extrapolations show that the large-scale field of AB Dor resembles a tilted dipole, giving (in addition to many small-scale bright regions) two dominant bright regions at opposite longitudes. The latitudes of these bright regions can be related (crudely) to the tilt of this large-scale dipole. Depending on the stellar inclination one or both may be visible. We therefore explore the impact on the X-ray emission from these T Tauri stars that a field with a realistic degree of complexity might have.

\section{Fields with a realistic degree of complexity}
Although the observations to date suggest that T Tauri stars have fields that are more complex than a simple dipole, we do not as yet have any direct observations of the detailed geometry of the field. We therefore use fields extrapolated from surface magnetograms of two young solar-like stars, AB Dor (period 0.514 days) and LQ Hya (period 1.6 days) that have been obtained using Zeeman-Doppler imaging \cite{donati97zdi,donati03,donati97abdor95,donati99abdor96}. This allows us to determine the influence of field complexity and to assess the effect that this might have on our results based on simple dipolar fields. Since the rotation axes of both of these stars are inclined to the observer, only one hemisphere can be imaged reliably. In order to provide a realistic surface map for the hidden hemisphere, we use a map taken from another year. Thus for the first map we use magnetograms of AB Dor generated from data acquired in December 1995 and December 1996, and for the second map we use magnetograms of LQ Hya generated from data acquired in December 1993 and December 2001. The coronal fields are extrapolated from these using the ``Potential Field Source Surface'' method originally developed by \scite{altschuler69} for extrapolating the Sun's coronal field from solar magnetograms. We use a code originally developed by \scite{vanballegooijen98}. Since the method has been described in \scite{jardine02structure} we provide only an outline here. Briefly, we
write the magnetic field $\bvec{B}$ in terms of a flux function $\Psi$
such that $\bvec{B} = -\bvec{\nabla} \Psi$ and the condition that the
field is potential ($\bvec{\nabla}\times\bvec{B} =0$) is satisfied
automatically.  The condition that the field is divergence-free then
reduces to Laplace's equation $\bvec{\nabla}^2 \Psi=0$ with solution 
in spherical co-ordinates $(r,\theta,\phi)$
\begin{equation}
 \Psi = \sum_{l=1}^{N}\sum_{m=-l}^{l} [a_{lm}r^l + b_{lm}r^{-(l+1)}]
         P_{lm}(\theta) e^{i m \phi},
\end{equation}
where the associated Legendre functions are denoted by $P_{lm}$.  The
coefficients $a_{lm}$ and $b_{lm}$ are determined by imposing the
radial field at the surface from the Zeeman-Doppler maps and by
assuming that at some height $R_s$ above the surface (known as the {\em source surface}) the field becomes
radial and hence $B_\theta (R_s) = 0$.  This second condition models
the effect of the plasma pressure in the corona pulling open field
lines to form a stellar wind.  
 \begin{figure*}
  \begin{center}
   \includegraphics[width=3in]{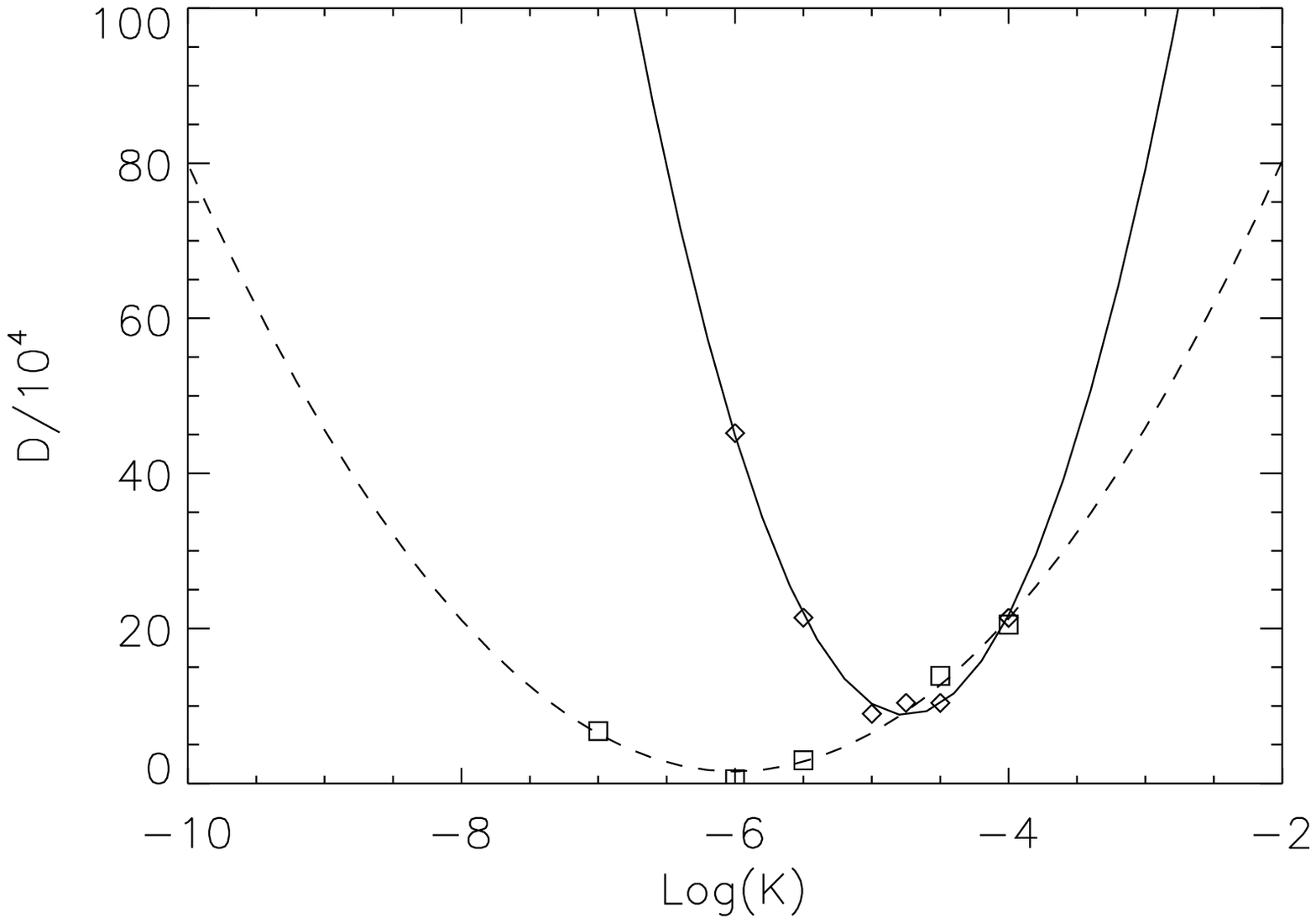}
   \includegraphics[width=3in]{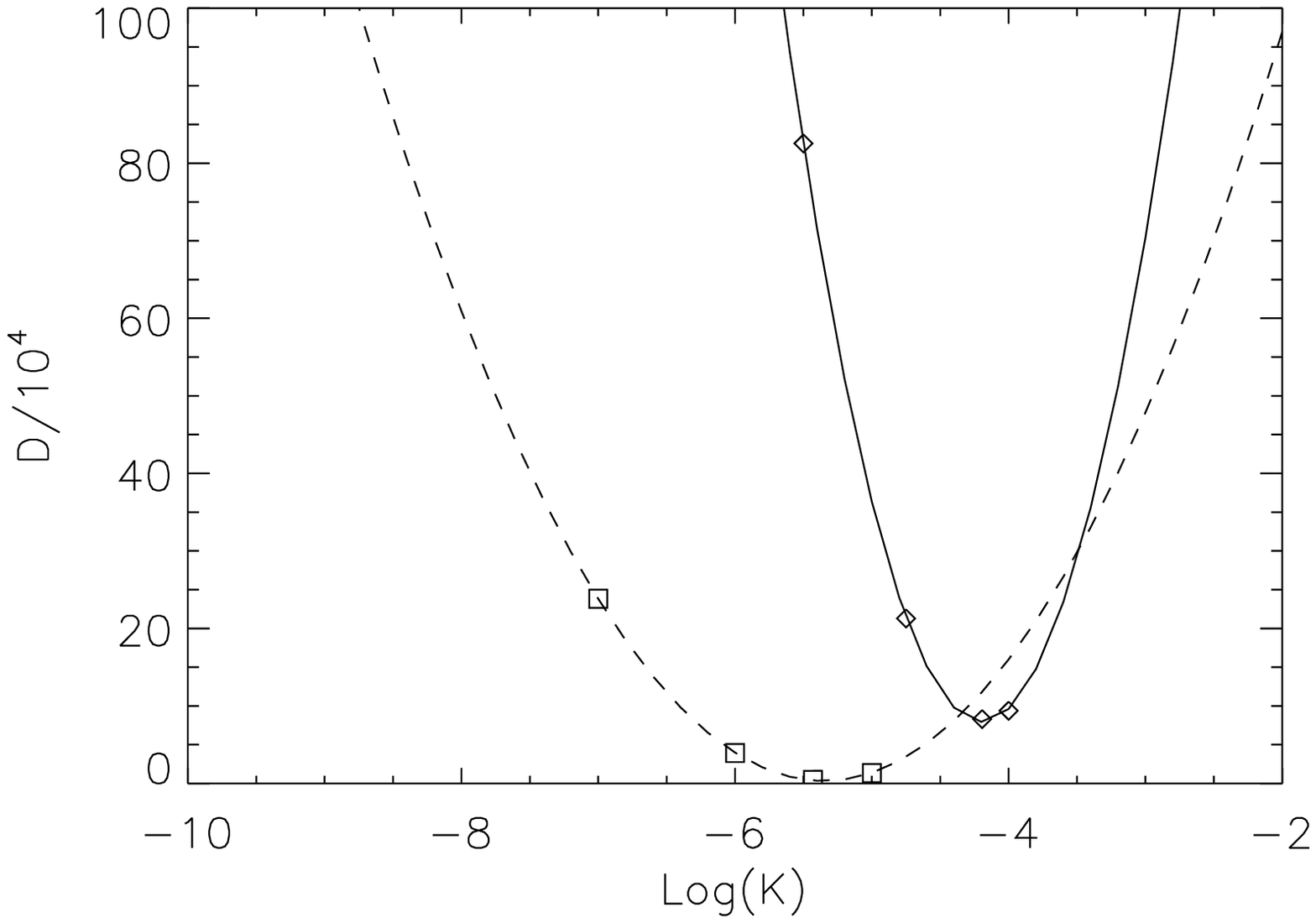}
   \caption{Variation of D (a measure of the goodness of fit to the data) with the base pressure for (left) the field structure shown in Fig. \ref{field_geom_mxd180}  and (right) the field structure shown in Fig. \ref{field_geom_lqhya}. The solid lines denote the higher-temperature coronae, and the dashed line the lower temperature coronae.}
   \label{min_chisq_complex}
 \end{center}
\end{figure*}

For rapidly-rotating solar-like stars, we can use the observed locations of the large slingshot prominences that typically form in their coronae to estimate the radii at which the corona must still be closed. For T Tauri stars, however, we do not have this option. As a fairly conservative estimate we choose to take the radial distance at which a {\em dipolar} field (of the same average field strength) would remain closed.  We choose the dipolar field as it gives the greatest coronal extent (since of all the possible multipolar components, the dipolar falls off most slowly with radial distance). If this maximum extent exceeds the co-rotation radius, we set it to be the co-rotation radius instead. 

We determine the pressure at every point by calculating the path of the field line through that point and solving hydrostatic equilibrium along that path. Once again, we set the gas pressure at the base of the field line to be proportional to the magnetic pressure there, noting however that now the base pressure varies across the stellar surface. By varying the constant of proportionality $K$ we can raise or lower the overall level of the gas pressure. If along any field line we find that the gas pressure exceeds the magnetic pressure, we assume that the field line should have been opened up by the gas pressure and so we set $p=0$.

We assume that the disk occupies a range of latitudes extending some $10^\circ$ above and below the equator. We do not set the location of the inner edge of the disk, but allow it to adjust locally to the structure of the magnetic field at each latitude and longitude. We therefore assume that the disk extends in towards the surface of the star until it reaches the first field line along which a) the gas pressure is less than the magnetic pressure everywhere along the field line and b) the net gravitational force at the point where the field line passes through the disk acts inward towards the star. We assume that this first field line carries all the accreting material from that particular latitude and longitude and that it shields any field lines interior to it.  Figs. \ref{field_geom_mxd180} and \ref{field_geom_lqhya} show the structure of these accreting field lines, examples of the closed field lines that could contain hot coronal gas and  X-ray images from the same perspective. 
 \begin{figure*}
  \begin{center}
   \includegraphics[width=3in]{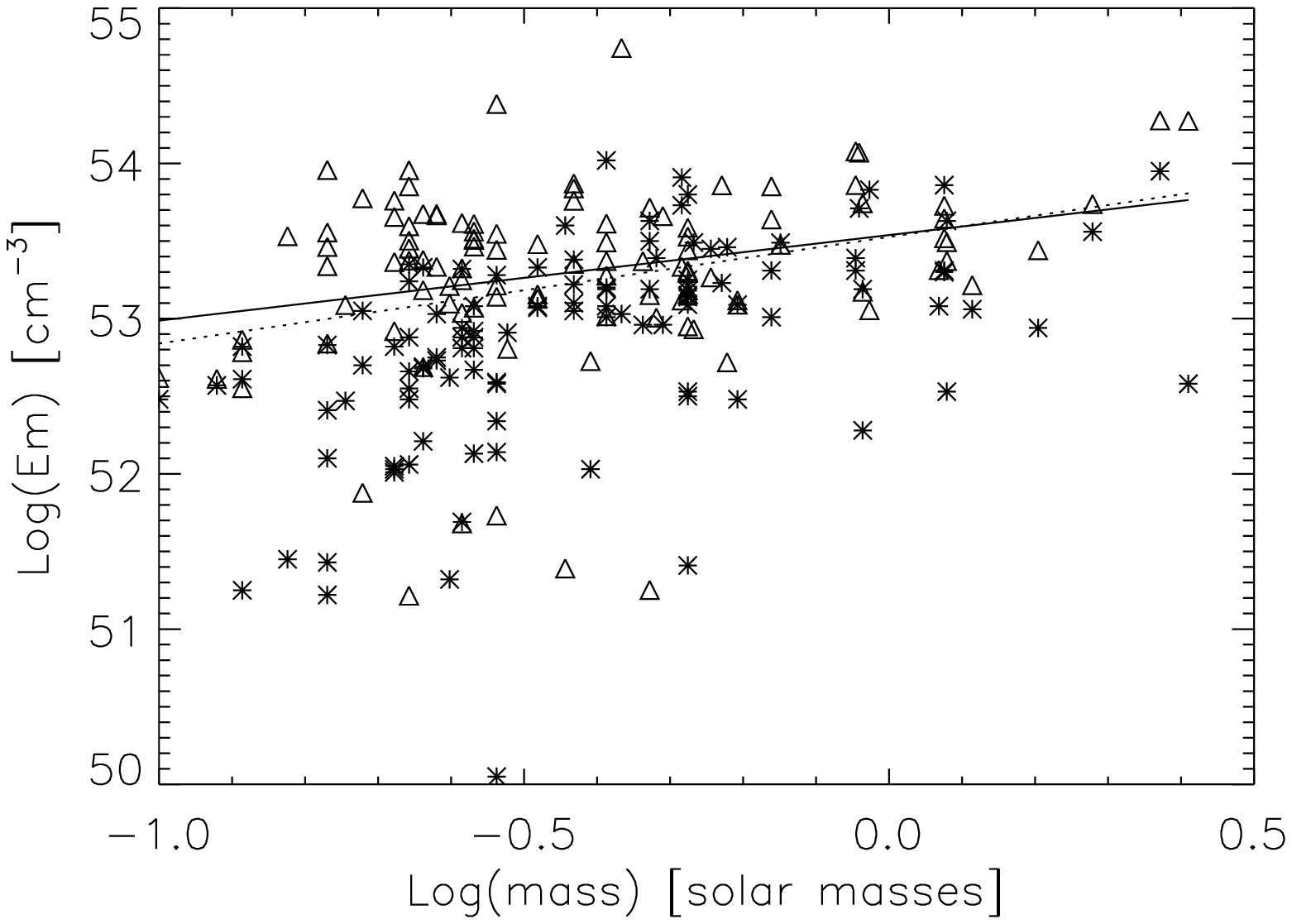}
   \includegraphics[width=3in]{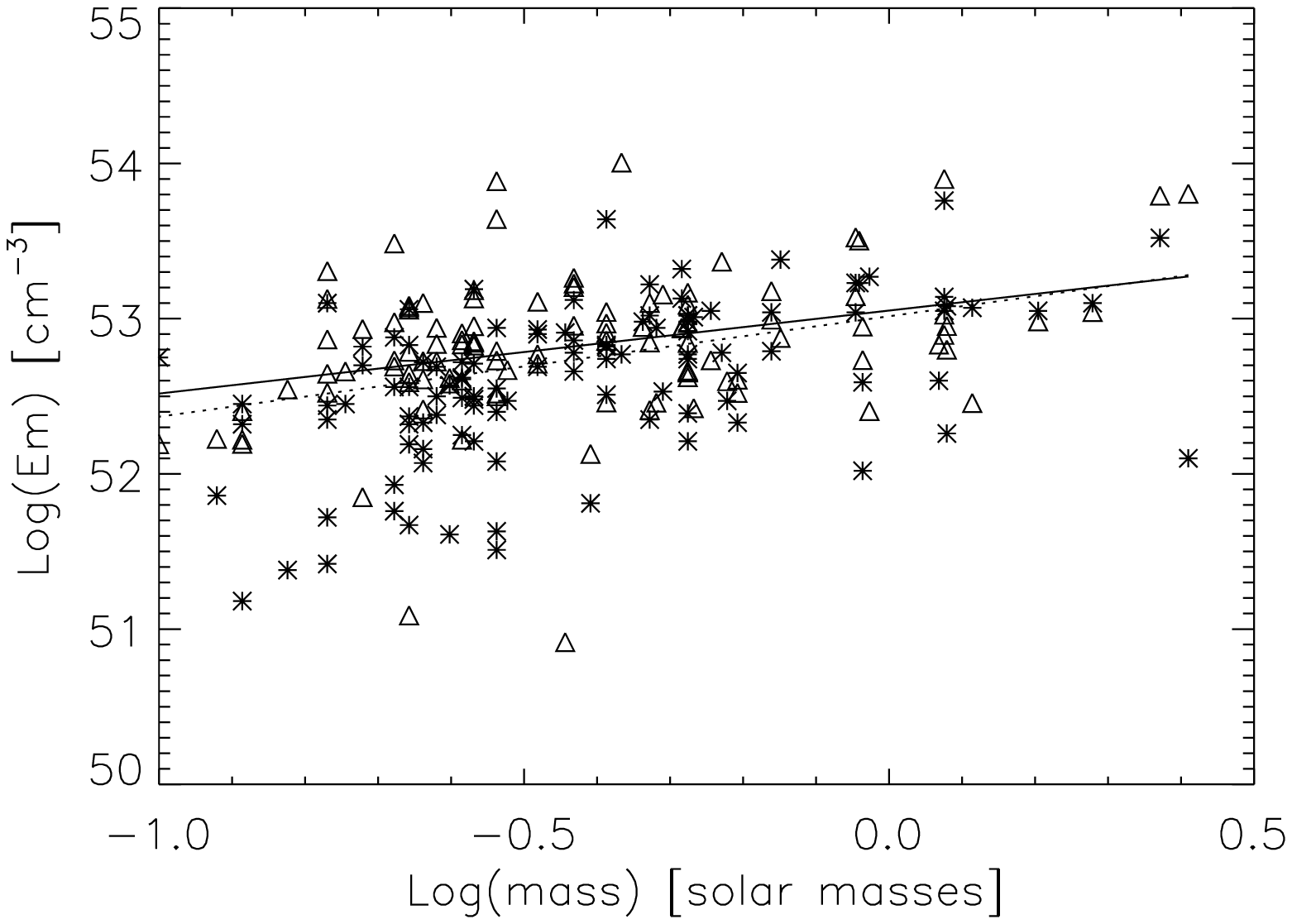}
   \includegraphics[width=3in]{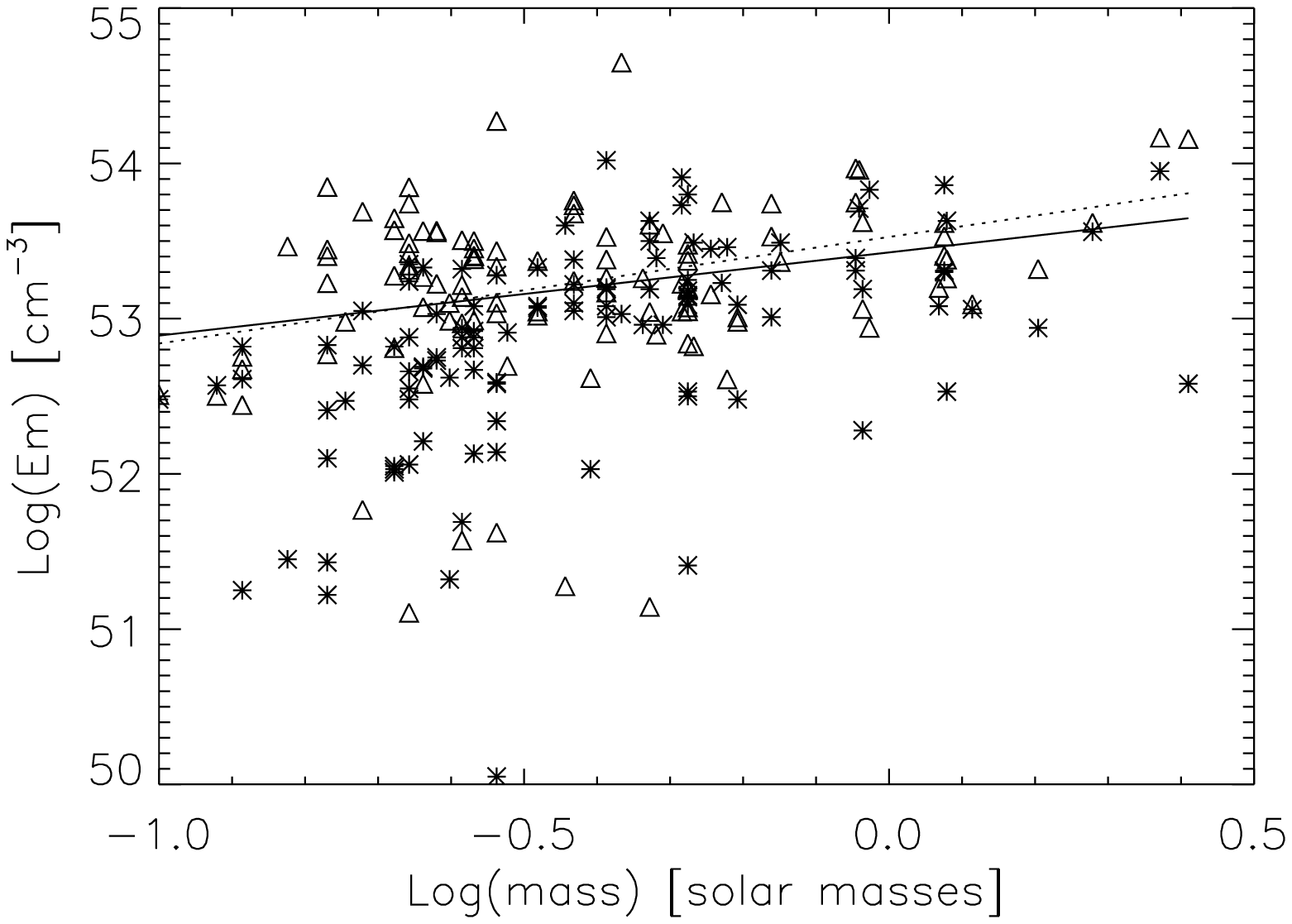}
   \includegraphics[width=3in]{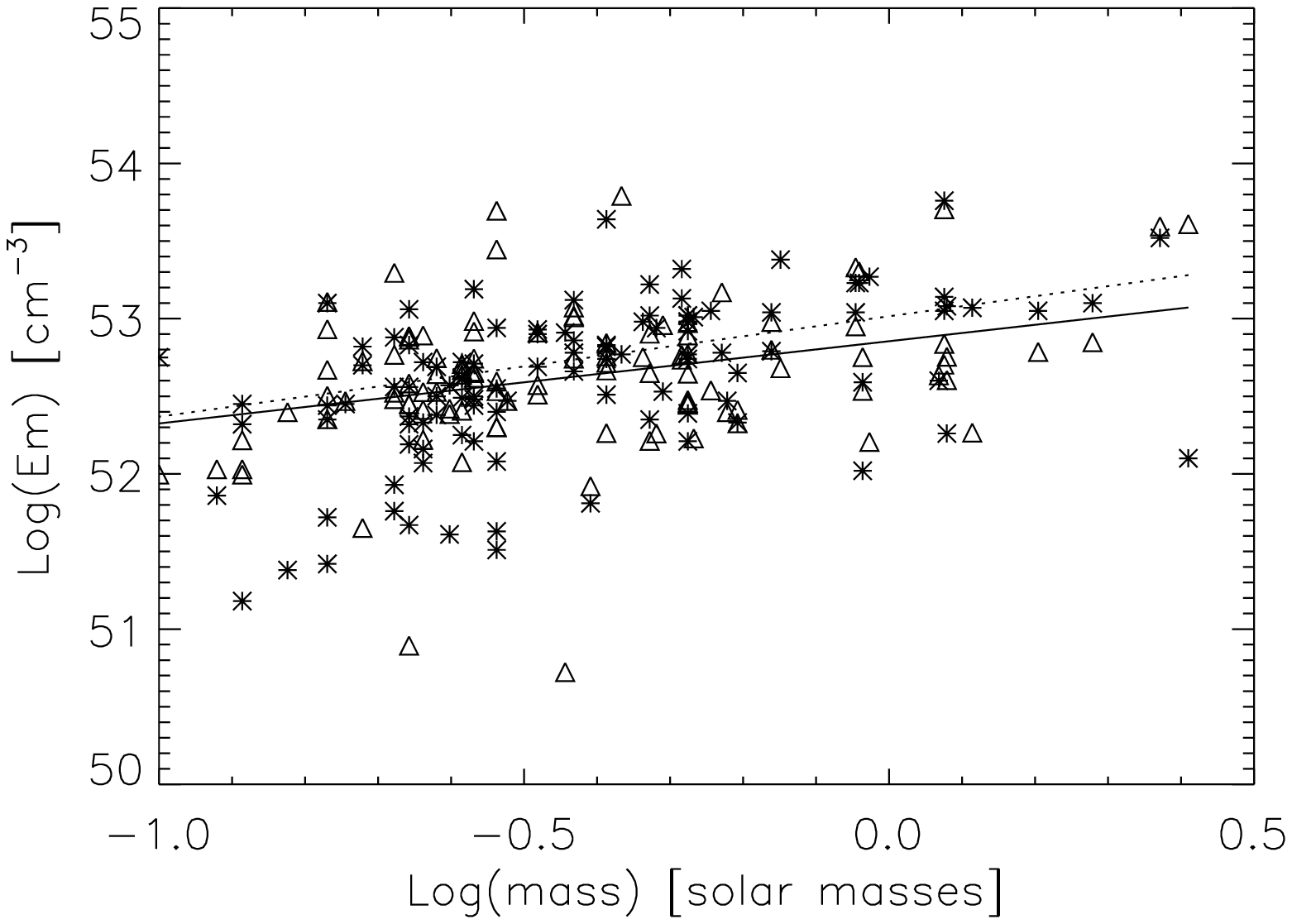}
   \caption{Observed emission measures for low (left) and high (right) temperature ranges from the COUP sample. The top panels refer to the field structure shown in Fig. \ref{field_geom_mxd180} while the bottom panels refer to the field structure shown in Fig. \ref{field_geom_lqhya}. The solid line is the best fit to the calculated emission measures, while the dotted line is the best fit to the observed emission measures.}
   \label{em_complex}
 \end{center}
\end{figure*}

Of our two free parameters, one (the flux density or field strength) is determined by the surface magnetograms. As before, the other parameter is the constant of proportionality $K$ that determines the gas pressure at the base of each field line. For the dipole field, we chose for simplicity to have a uniform $p_0$ based on the average field strength $<|B|>$. For these complex fields, however, the base magnetic pressure can vary significantly across the surface and as a result, so can the base gas pressure. In Table \ref{table2} therefore it should be noted that the value of $p_0 = K<|B|>^2$ is just the value based on the average value of $<|B|>$.

We can therefore, for each of the two example field structures, determine which value of K  gives the best fit to the emission measures from the COUP survey (see Fig. \ref{min_chisq_complex}). We show in Fig. \ref{em_complex} the calculated emission measures superimposed on the observed emission measures. As can be seen from Table \ref{table2} the slopes for these calculated values are shallower than for the observed emission measures, but the values for the two field structures are remarkably similar.

There is more of a variation to be found in the densities. For the dipole field, densities of around $0.4\times 10^{10}$cm$^{-3}$ are typical for the high temperature range. For these more complex fields, however, the densities are almost an order of magnitude higher ($2.5\times10^{10}$cm$^{-3}$). The slopes are also less than for the active disk with a dipolar field. This is probably because the more compact fields can support a higher density, but they do not extend out as far as the dipolar fields and so they are less affected by the presence of a disk.
 \begin{table}
\caption{This shows, for two example field structures with a realistic degree of complexity, the average field strength $B_0 = <|B|>$  at the base of the corona and the corresponding gas pressure $p_0 = KB_0^2$. By fitting relationships of the form $EM \propto M_\star^a$ and $\bar{n}_e \propto M_\star^b$ to the calculated values of the emission measures and densities we have determined the values of $a$ and $b$.}
\begin{center}
\begin{tabular}{|c|c|c|c|c|}
\hline
               & AB Dor-like &            &   LQ Hya-like     &    \\
\hline
  T             &High         & Low   & High              & Low  \\
\hline
 $<|B|> [G]$ &143         & 143   & 143               &  143 \\
 \hline
   $p_0$ [dyne cm$^{-2}$] & 3.6           & 0.2     & 13.2                & 0.9 \\
 \hline
 a            &0.55           & 0.53     & 0.54                & 0.53 \\
 \hline
 b           &-0.05          & -0.11   & -0.06                   & -0.12\\
 \hline

\end{tabular}
\end{center}
\label{table2}
\end{table}
 \begin{figure*}
\begin{center}
  \includegraphics[width=3in]{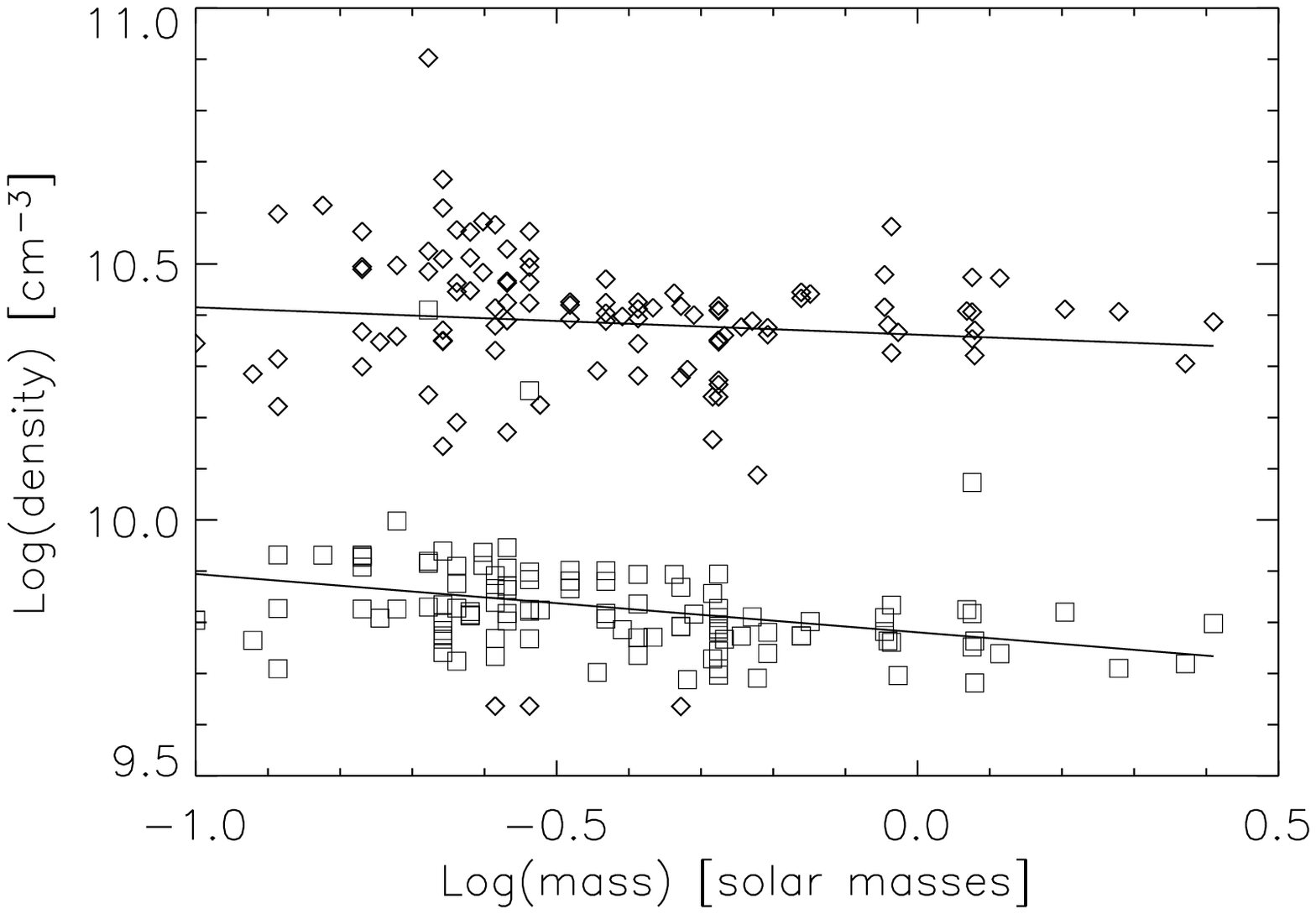}
    \includegraphics[width=3in]{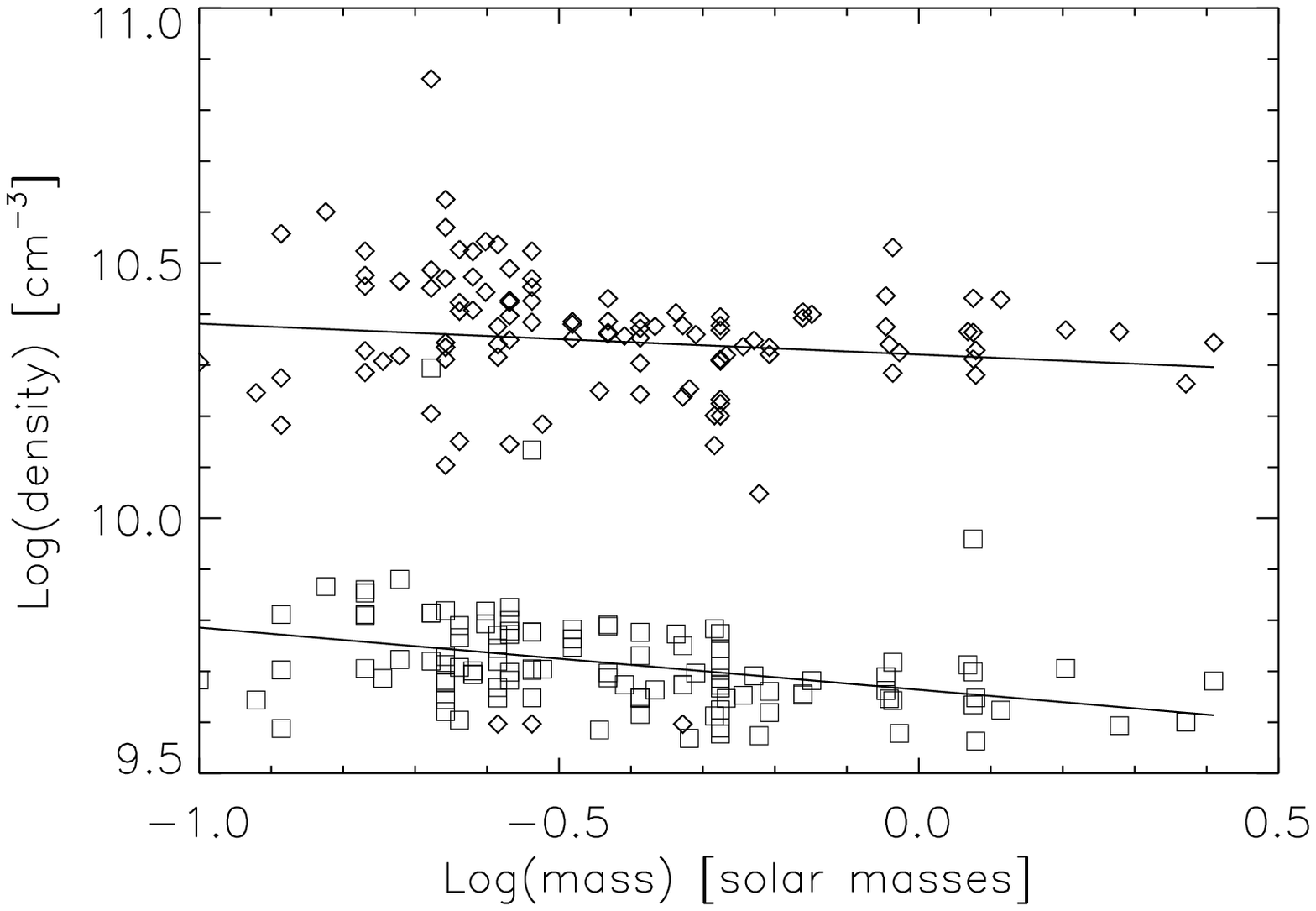}
  \caption{Emission-measure weighted densities for both the high temperature (diamonds) and low temperature (squares) ranges from the COUP sample. The left panel refers to the magnetic fields shown in Fig. \ref{field_geom_mxd180} and the right panel to the magnetic fields shown in  Fig. \ref{field_geom_lqhya}.}
  \label{dens_complex}
 \end{center}
\end{figure*}
 \begin{figure*}
\begin{center}
  \includegraphics[width=3in]{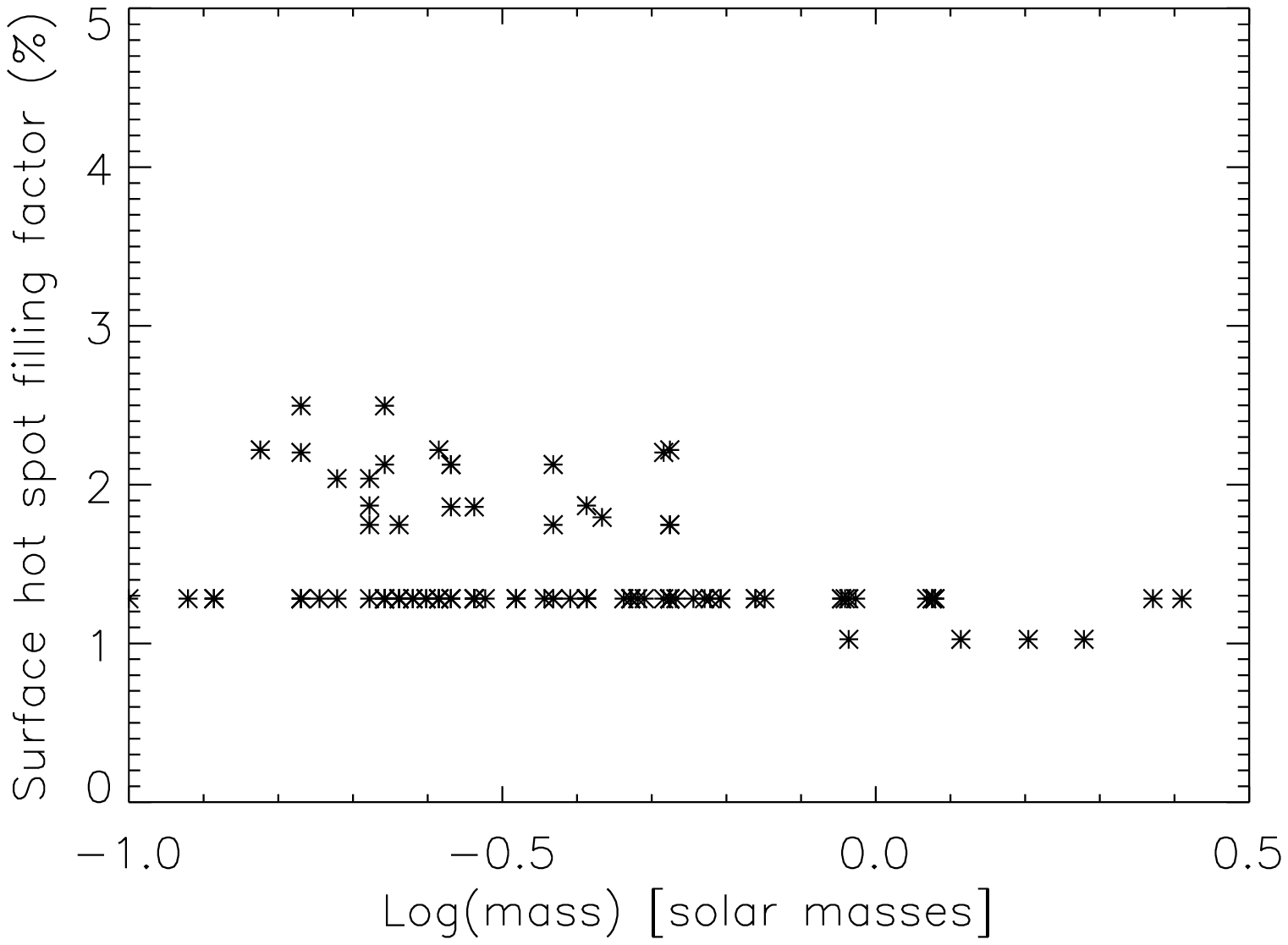}
  \includegraphics[width=3in]{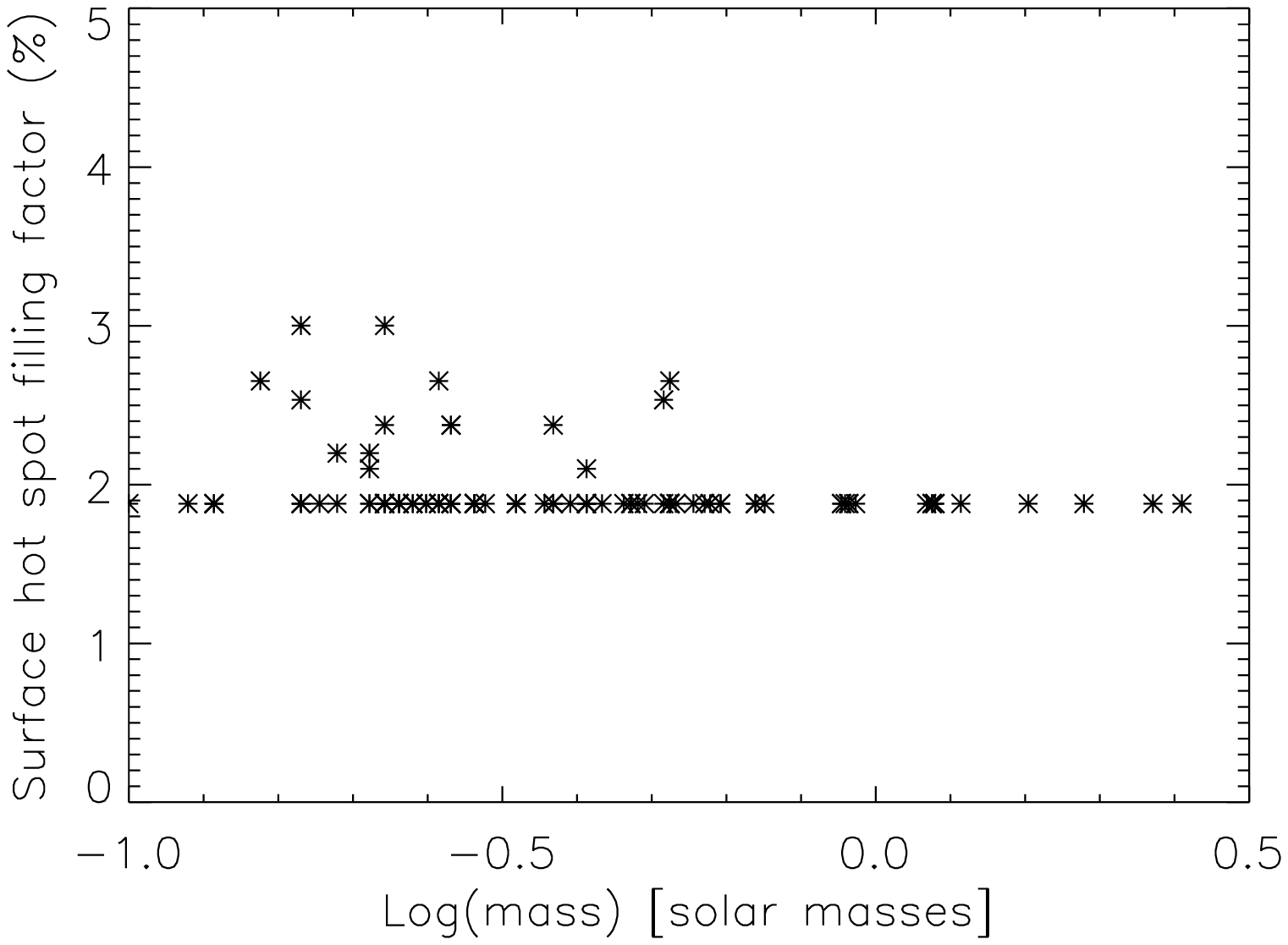}
  \caption{Surface filling factor of hotspots for  (left) the magnetic fields shown in Fig. \ref{field_geom_mxd180} and (right) the magnetic fields shown in Fig. \ref{field_geom_lqhya}.}
  \label{hotspots}
 \end{center}
\end{figure*}

 
 With these more complex fields we would expect to find hotspots that are in different surface locations to those for a dipole field and that are perhaps different in size.  Fig. \ref{hotspots} shows the fraction of the stellar surface area that is covered in field lines that pass through the inner edge of the disk.  For the higher mass stars whose coronae do not reach out to the disk, the field lines that reach the disk are open and radial. Variations in the position of the co-rotation radius then have little effect in determining which field lines pass through the disk and so the fraction of accreting field lines varies little with stellar mass. For those lower-mass stars whose coronae extend out beyond the co-rotation radius, the stellar magnetic field is still closed at the inner edge of the disk. Consequently, changes in the co-rotation radius can make large changes to whether a field line is accreting or not and so  the surface area of hot spots shows a greater scatter. In either instance, however, the filling factor is low, in agreement with results from modelling of accretion signatures \cite{muzerolle_mdot_03,calvet_highmass_04,muzerolle_vlm_05,valenti_johns_krull_04,symington_tts_05}.

 \section{Summary and Discussion}
 
 We have shown in this paper that it is possible for a simple model to reproduce the observed variation with mass of the X-ray emission measure of T Tauri stars. This variation is due to the restriction in the size of the stellar corona that results either from the pressure of the hot coronal gas in opening up the magnetic field lines, or from the shearing effect of a circumstellar disk.  The ratio of the pressure scale height to the co-rotation radius (a reasonable estimate of the location of the inner edge of an accretion disk) scales as $\Lambda / R_{\rm KCR} \propto M_\star^{-2}$ for a polytropic star. Hence, lower mass stars are more likely to have coronae that reach the inner edge of any accretion disk present and so their coronal extent will be set by the co-rotation radius. Higher mass stars have coronae that may not extend as far as the co-rotation radius and so their coronal extent is set simply by the ability of the magnetic field to contain the gas. For these stars, accretion will occur principally along open, nearly radial magnetic field lines.
 
 Even in the absence of a disk, the variation of the extent of the corona will produce a drop in the X-ray emission measure with decreasing stellar mass. This process of {\em coronal stripping} has already been shown to reproduce the observed variation with rotation rate of both the magnitude and rotational modulation of the X-ray emission measures of main-sequence stars \cite{unruh97loops,jardine99stripping,jardine04stripping}. The stars in the COUP sample appear to lie in the saturated or supersaturated part of the rotation-activity relation (i.e. where  X-ray emission measure is independent of rotation rate or even falling with increasing rotation rate). Their emission measures are slightly lower however than their main-sequence counterparts and they show a greater scatter. \cite{preibisch_COUP_insights_05} have examined this question of the overall suppression of the X-ray emission. They divide the sample of stars into those that show some indication of the presence of a disk (i.e. an infra-red excess) but no apparent signatures of active accretion (as measured by the width of the H$_\alpha$ line) and those that show indications of both. The main contributors to the drop in X-ray emission appear to be those stars that show signs of active accretion and for which there is therefore clear evidence that their disks are interacting with the stellar magnetic field. Stars lacking this clear signature of disk interaction appear to show a level of X-ray emission that is similar to their saturated or supersaturated main sequence counterparts. This is consistent with our model which predicts that all stars will show a drop in emission with decreasing mass due to the pressure stripping of their coronae, but that in those stars where a disk is actively interacting with and distorting the stellar magnetic field, this drop in emission will be further enhanced. \scite{preibisch_COUP_insights_05} also note that the large scatter in X-ray emission is most pronounced in the actively accreting stars. This can be understood in terms of our model since in those stars where the disk is actively stripping the corona, the coronal extent depends on a further parameter, the rotation rate, since this determines the location of the co-rotation radius where the inner edge of the disk is most likely to be situated.
  
The observations of the strength of the magnetic field at the base of the accretion funnels and across the rest of the stellar surface \cite{valenti_johns_krull_04,symington_tts_05} combined with the observation of significant X-ray rotational modulation in  stars in the COUP sample \cite{flaccomio_COUP_rotmod_05} have prompted us to investigate the importance of the nature of the stellar magnetic field. We have modelled the extent of the stellar corona, the X-ray emission measure and the density both for a simple dipole field and also for fields with a realistic degree of complexity. We do not yet know in detail the structure of T Tauri coronal fields, but we have used instead coronal magnetic fields extrapolated from surface magnetograms of young main-sequence stars. While we do not yet know if these capture the coronal structure of T Tauri stars, they allow us to explore the effect of a more compact field. This is important as dipole fields (the most commonly used example until now) fall off with height more slowly than any other type of field except a purely radial one. They therefore model the most extended type of field possible. Fields that are more complex are more compact: their field strengths decay with height more rapidly than a dipolar field and so they will typically be less extended than a dipole and consequently are likely to produce a higher density.

We find that the more complex fields do indeed produce more compact, higher density coronae, with densities in the range $(2.5-0.6)\times 10^{10}$cm$^{-3}$ for coronae whose temperatures are set to the ``high'' and ``low'' temperatures in the COUP database. These densities are nearly an order of magnitude higher than for a dipole field. The densities produced by different types of complex fields are relatively similar, with the main differences being the the locations of the surface hotspots (where the accretion flows impact on the stellar surface) and their filling factors. In contrast to dipole fields where any accretion flow must reach the surface at the poles (thereby producing no rotational modulation), more complex fields can produce hotspots at lower latitudes. \scite{mahdavi98} have modelled the locations of hotspots in dipole field that are tilted with respect to the rotation axis, but these can only provide one hotspot in each hemisphere, while more complex fields can produce many smaller hotspots at a range of latitudes and longitudes. We find filling factors are small (typically a few percent) in agreement with predictions made from modelling accretion signatures \cite{muzerolle_mdot_03,calvet_highmass_04,muzerolle_vlm_05,valenti_johns_krull_04,symington_tts_05}.

 We have also calculated the form of the relationship $EM \propto M_\star^a$ based on the published stellar masses, radii, rotation rates and coronal temperatures for the COUP stars \cite{getman_COUP_list_05a,getman_COUP_list_05b}. For the higher temperature range in the COUP sample, the observed emission measures (allowing for the published errors) give a value of $a = 0.68$. If we assume that all stars have either no disk present, or only a {\em passive} disk that does not distort the stellar corona, we find a value of $a=0.34$ for a dipole field, compared with $a=0.70$ if we assume that all stars have an {\em active} disk that distorts any stellar corona that extends as far as the inner edge of the disk (which is assumed to be at the co-rotation radius).  If we assume a more complex field geometry this value drops to $a=0.55$ if the stellar disks are assumed to be active. A similar result is found for the low temperature component. Thus the presence of disks capable of distorting those stellar coronae that extend beyond the co-rotation radius seems to be necessary to explain the slope of the observed variation with mass of the X-ray emission measure. The degree of complexity of the field has a smaller effect on the overall X-ray emission, but it does have a significant effect on the densities. Assuming that all stars have dipole fields gives densities much lower than those derived from observations of flares \cite{favata_COUP_flares_05}, while the assumption of a more complex field geometry gives densities comparable to the published values. These results taken together suggest that the true T Tauri stellar fields may be more extended than those of the main sequence counterparts, but not by as much as a purely dipole field. They may simply have a dipolar component that is somewhat stronger than in the main sequence stars.
 
 In summary, we have shown that a simple model (which reproduces the X-ray activity- rotation relation for main sequence stars) can explain the variation of X-ray emission measure with stellar mass observed in the COUP survey, if we allow for the presence of an accretion disk which will limit the size of any stellar corona that reaches out to its inner edge.
\section{Acknowledgements}

The authors would like to thank Aad van Ballegooijen who wrote the original version of the field extrapolation software. 

\section{appendix}
In deriving expressions for the variation of the emission measure and density with radial distance from the star, we begin with (\ref{grav}). If we scale all distances to a stellar radius, we can write
\begin{equation}
 p(r,\theta) = p_\star(1,\theta)\exp\left[ \Phi_g(\frac{1}{r} -1) + \Phi_c\sin^2\theta(r^2-1)\right]
\end{equation}
where $p_\star(1,\theta)$ is the pressure distribution across the stellar surface and 
\begin{eqnarray}
 \Phi_g & = & \frac{GM/r_\star}{k_BT/m} \\
 \Phi_c & = & \frac{\omega^2r_\star^2/2}{k_BT/m}
\end{eqnarray}
are the surface ratios of gravitational and centrifugal energies to the thermal energy. The emission measure $Em = \int n_e^2 dv$ is then
\begin{eqnarray}
 EM & = & \frac{4\pi r_\star^2}{(k_B T)^2} \\ \nonumber
        &    & \int_{\theta = \Theta_m}^{\theta = \pi/2}
       p_\star^2(1,\theta)\exp \left [-2\Phi_c \cos^2 \theta (r^2-1) \sin\theta d\theta \right] \\ \nonumber
        &    &  \int_{r=1}^{r=r_m}r^2\exp \left[2\Phi_g (\frac{1}{r}-1 + 2 \Phi_c(r^2 -1) \right] dr. \nonumber
\end{eqnarray}
If we now use the substitution $\mu = \cos \theta$ such that $\mu_m = \cos \Theta_m = (1-r/r_m)^{1/2}$ then the $\theta$-integral can be written as 
\begin{equation}
 \int_{0}^{\mu_m}p_\star^2(1,\theta)\exp \left[ -2\Phi_c \mu^2 (r^2-1) \right] d\mu
\end{equation}
which, on using the substitution $t=[2\Phi_c (r^2-1)]^{1/2} \mu $ gives
\begin{equation}
 \frac{1}{2\Phi_c (r^2-1)^{1/2}} \int_0^{t_m} p_\star^2(1,\theta) e^{-t^2} dt
\end{equation}
where $t_m = [2\Phi_c (r^2-1)]^{1/2} [1-r/r_m]^{1/2}$. If we now assume that the pressure is uniform over the stellar surface we obtain the following expression for the emission measure as a function of $r$:
\begin{eqnarray}
  EM(r) & = & \sqrt{ \frac{2\pi^3} {\Phi_c} }\frac{r_\star^3 p_\star^2}{(k_B T)^2} \\ \nonumber
        &    &\int_{r=1}^{r=r_m}
               \frac{r^2}{(r^2 - 1)^{1/2}} {\rm erf} \left[2\Phi_c(r^2-1)(1-\frac{r}{r_m}) \right]^{1/2}\\ \nonumber
         &  &      \exp\left[2\Phi_g  (\frac{1}{r}-1) 2 \Phi_c(r^2 -1)\right] dr. \\ \nonumber
\end{eqnarray}
We can, in a similar way, obtain an expression for the emission measure weighted density $\bar{n}_e(r) = \int n_e^3 dV / \int n_e^2 dV$. \begin{eqnarray}
\bar{n}_e(r) & = & \frac{1}{EM(r)}\sqrt{ \frac{\pi^3} {3 \Phi_c} }\frac{2 r_\star^3 p_\star^3}{(k_B T)^3} \\ \nonumber
        &    &\int_{r=1}^{r=r_m}
               \frac{r^2}{(r^2 - 1)^{1/2}} {\rm erf} \left[3\Phi_c(r^2-1)(1-\frac{r}{r_m}) \right]^{1/2}\\ \nonumber
         &  &      \exp\left[3\Phi_g  (\frac{1}{r}-1) 3 \Phi_c(r^2 -1)\right] dr \\ \nonumber
\end{eqnarray}
where we remind the reader that all distances $r$ have been scaled to a stellar radius.




\end{document}